\documentclass[11pt, oneside]{article}   	
\usepackage{geometry}                		
\usepackage{graphicx}				
\usepackage{amssymb}

\usepackage{newtxtext}
\usepackage{newtxmath}
\usepackage{natbib}
\usepackage{hyperref}
\usepackage{epstopdf, epsfig}
\usepackage[utf8]{inputenc}
\usepackage{amsmath}
\usepackage{color}
\usepackage{listings}
\usepackage{float}
\usepackage{xcolor}
\usepackage{placeins}
\hypersetup{
    colorlinks = false,
    urlcolor   = blue,
    citecolor  = black,
}

\newcommand{\RomanNumeralCaps}[1]
\linenumbers
\title{\textbf{Compositional noise in nozzles with dissipation}}
\author{Animesh Jain$^1$ \& Luca Magri$^{2,1}$\footnote{l.magri@imperial.ac.uk} \\ 
 \small{$^1$ Department of Engineering, University of Cambridge,
Cambridge CB2 1PZ, UK} \\
\small{$^2$ Imperial College London, Aeronautics Department, London, UK}}

\date{}

\begin{document}
\maketitle

\begin{abstract}
We propose a physical model to predict indirect noise generated by the acceleration of compositional inhomogeneities in nozzles with viscous dissipation (non-isentropic nozzles). 
First, we derive the quasi-one-dimensional equations from the conservation laws of multicomponent flows. 
Second, we validate the proposed model with the experimental data available in the literature for binary mixtures of four gases.  
Third, we calculate the nozzle transfer functions for different Helmholtz numbers and friction factors, in both subsonic  and supersonic flows with/without shock waves. We show that friction and dissipation have a significant effect on the generation of indirect noise, for which the physical mechanism is identified and explained.
Fourth, we find a semi-analytical solution with path integrals, which provide an asymptotic expansion with respect to the Helmholtz number. 
Fifth, we introduce the compositional noise scaling factor, which is applied to quickly estimate compositional noise from the knowledge of only one single-component gas transfer function. The approximation error is less than $1\%$. 
The proposed low-order model provides accurate estimates of the transfer functions and physical insight into  indirect noise for multicomponent gases. This opens up new possibilities to accurately predict, and understand, sound generation in gas turbines. 
\end{abstract}

\section{Introduction}
As aeroengines and gas turbines become cleaner, the combustion process is becoming a major source of noise emissions. This is because the reduction of air pollutant emissions is achieved with lean flames, which, in turn, burn unsteadily to generate sound waves through direct and indirect physical mechanisms.
On the one hand, the sound generated by the unsteady heat released by the flame, which leads to a volumetric contraction and expansion of the gas, is referred to as direct combustion noise~\citep[e.g.,][]{mahmoudi2018low, ihme2017combustion}. On the other hand, the sound generated by the acceleration of flow inhomogeneities through the nozzles downstream of the combustor is referred to as indirect noise \citep[e.g.,][]{marble1977acoustic, strahle1976noise, cumpsty1979jet, williams1975generation, polifke2001constructive, morgans2016entropy, magri2016compositional}. 
These sound waves may have two detrimental effects:
(i) they contribute to noise pollution; and 
(ii) they can reflect back in the combustion chamber and, if they are sufficiently in phase with the heat released by the flame, can cause thermoacoustic oscillations to arise \citep[e.g.,][]{polifke2001constructive, goh2013influence, motheau2014mixed}. 
Indirect noise caused by temperature inhomogeneities is referred to as entropy noise \citep{cuadra1967acoustic, marble1977acoustic, bake2009entropy, duran2013solution},  whereas indirect noise caused by compositional inhomogeneities is referred to as compositional noise~\citep{magri2016compositional, magri2017indirect}. The third type of indirect noise is vorticity noise, which is generated by velocity gradients  {\citep{howe1975contributions, hirschberg2021sound, hirschberg2022sound}}. In aircraft engine applications, however, the role of vorticity noise is typically negligible \citep{dowling2015combustion}. 
Although there are many studies on direct noise, which make it a relatively well-understood mechanism \citep{ihme2017combustion}, a complete understanding of indirect noise is yet to be developed \citep{tam2019combustion}. 
Experimentally, isolating the effect of indirect noise requires pressurised experimental rigs and close-to-anechoic boundary conditions, which make the design of experimental campaigns challenging \citep{rolland2018sound}. In this paper, we focus on entropy and compositional noise.\\

Most of the studies in the literature consider the flow to consist of a single component. However, due to factors such as dilution and imperfect mixing, the exhaust gas has fluctuations in the mixture composition \citep{magri2016compositional, magri2017indirect}. To model multicomponent gases, the compact nozzle assumption for single-component flows developed by \citet{marble1977acoustic} was extended to account for impinging compositional waves by \citet{magri2016compositional}, who showed the role of compositional inhomogeneities in indirect noise generation in subsonic and supersonic compact nozzles.  These studies recovered the entropic-acoustic and acoustic-acoustic transfer functions obtained by \citet{marble1977acoustic} in the limit of a  homogeneous mixture. 
The compact nozzle assumption and theory of compositional noise were generalized by \citet{magri2017indirect}, who derived the  governing differential equations, and identified the physical sources in finite-length nozzles and impinging waves with non-zero frequency. They found an expression for the  density as a function of key thermo-chemical parameters, which were shown to be significant dipole sources of sound. The frequency-dependent behaviour of compositional noise was analysed in subsonic and supersonic nozzles.  Compositional noise was shown to monotonically decrease with the Helmholtz number with a semi-analytical solution of the governing equations with a Dyson expansion, as reviewed in~\citet{magri2023linear}. For aeronautical applications, compositional noise was shown to be as great as, or larger than, entropy noise for a  kerosene mixture in a supersonic regime~\citep{magri2018effects}. Similarly to the case of entropy noise, the compact nozzle overpredicted compositional noise. Compositional-noise sources in a  rich-quench-lean combustor were computed by a high-fidelity large-eddy simulation in a realistic aeronautical gas turbine \citet{giusti2019flow}. They found that compositional noise can have the same order of magnitude as entropy noise. 
{
Recently, different studies proposed models for nonlinear effects \citep{huet2013nonlinear}, multi-stream nozzles \citep{younes2019indirect}, heat transfer \citep{yeddula2022solution}, and three-dimensional effects of the entropy field \citep{emmanuelli2020description, huet2020entropy, yeddula2022magnus}.
}
{
In this work, we assume the acoustics to be dominated by quasi-one dimensional  dynamics.
}
Common to the aforementioned studies on compositional noise is the assumption that the flow is isentropic. In reality, the energy dissipation due to viscosity and wall friction makes the flow non-isentropic. With the compact nozzle assumption and the heuristic argument of~\citet{de2019generalised}, \citet{rolland2018sound, de2021compositional} derived the nozzle transfer functions with appropriate jump conditions for non-isentropic multicomponent flows. They performed the analysis on a subsonic flow through a compact nozzle. 
\citet{rolland2018direct} experimentally studied the compositional inhomogeneities of air-helium mixtures that are accelerated through choked compact nozzles, for which a mass injection device was employed to validate the model. Recently, the effect of non-isentropicity was validated experimentally by injecting pockets of argon, carbon dioxide, helium and methane accelerating through isentropic nozzles by \citet{de2021compositional}. 
 The results were compared with the transfer functions obtained with a low-order physics-based model for non-isentropic nozzle flows, which relies on a semi-empirical parameter~\citep{de2019generalised}. 
 {
 \citet{huet2021influence} studied the influence of viscosity on entropy noise generation and scattering.
 Additionally, \citet{yang2020sound} and \citet{guzman2022scattering} modelled the effect of non-isentropicity in entropy noise generation in a sudden area expansion, which is a canonical flow with separation and
mean pressure losses. 
 }
 Recently, \citet{jain_magri_2022} derived the governing differential equations from first principles to model the non-isentropicity of nozzles with a spatial extent, which provided physical interpretation of  the indirect sound-generation process in non-isentropic nozzles. 
A detailed parametric study of the nozzle and flow parameters was carried out by \citet{jain2022sound}. 
 However, the physics was modelled only for a single component flow. \\

The overarching objective of this paper is to derive the equations from  first principles to model multicomponent flows in nozzles with viscous dissipation. 
Specifically, the goals are to 
(i) propose a physical model for a non-isentropic multicomponent nozzle flow from conservation laws;
(ii) investigate  compositional noise in subsonic and supersonic nozzle flows with/without shock waves;
(iii) derive a semi-analytical solution to calculate the transfer functions via an asymptotic expansion; 
{(iv)} introduce the compositional noise scaling  factor to quickly estimate compositional-acoustic transfer functions from single-component gases.
%
%
For this, a converging-diverging nozzle is numerically investigated. The results are compared with the experiments of \citet{de2021compositional}, where available. 
The paper is structured as follows. 
Section~\ref{sec:mathmodel} introduces the mathematical model.
%
%
Sections~\ref{sec:INtransferfunct_sub} and ~\ref{sec:INtransferfunct_sup} show the nozzle response for the subsonic and supersonic regimes, respectively. 
Section~\ref{sec:asymptexpn} presents the semi-analytical solution.
%
%
Section~\ref{sec:barK} introduces the scaling factor. 
%
%
Conclusions end the paper.


\section{Mathematical Model}\label{sec:mathmodel}

\begin{figure}
\centering    
\includegraphics[width=0.8\textwidth]{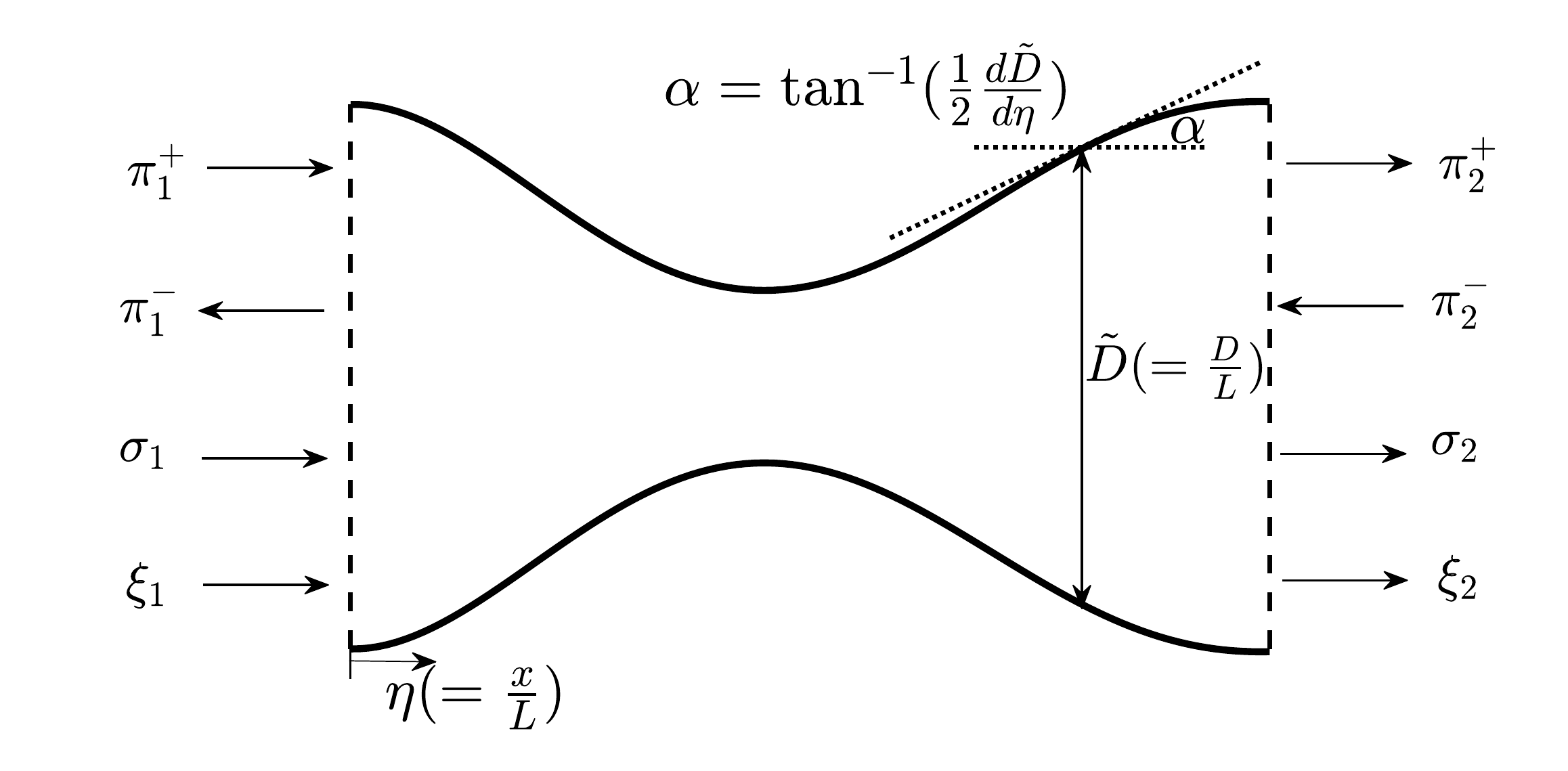}
\caption{Nozzle schematic  nomenclature.
}
\label{fig:nozzle_schem}
\end{figure}

The multicomponent gas mixture that accelerates through a nozzle is modelled under the following assumptions: 
(i) The flow is quasi-one-dimensional, i.e., the area variation causes a change in the flow variables, but the variables depend only on the axial coordinate;
(ii) the gas mixture consists of $N$ species of $Y_i$ mass fractions and $\mu_i$ chemical potentials, $\mu_i = W_i{\partial h}/{\partial Y_i} = W_i{\partial g}/{\partial Y_i}$, where $W_i$ is the molar mass of the $i$th species, $g$ is the specific Gibbs energy, and $h = \sum_{i=1}^N h_{i} Y_i$ is the specific enthalpy. The gas composition is parameterized with the mixture fraction, $Z$, as $Y_i = Y_i(Z)$ \citep{williams2018combustion}; 
(iii) the gases are assumed to be ideal with heat capacity $c_p = \sum_{i=1}^N c_{p,i} Y_i$, where $c_{p,i}$ is the heat capacity of the $i$th species at constant pressure; 
(iv) the gases are assumed to be calorically perfect, i.e., $c_{p,i}$ is constant, hence, the enthalpy is $h = c_p(T - T^o)$, where $T$ is the temperature and $^o$ is the reference state;
(v) the flow is considered to be chemically frozen; and 
(vi) the walls are adiabatic. 
With these assumptions, the equations of conservation of mass, momentum, energy and species are, respectively \citep{chiu1974theory}

\begin{align}
\frac{D\rho}{Dt} + \rho\frac{\partial u}{\partial x} + \frac{\rho u}{A}\frac{dA}{dx} &= {\Dot{S}_m}, \label{eq:dmdt_1}\\
    \frac{Du}{Dt} + \frac{1}{\rho}\frac{\partial p}{\partial x} &=  {\Dot{S}_M}, \label{eq:dmdt_132412}\\
    T\frac{Ds}{Dt} &= {\Dot{S}_s},\label{eq:dsdt_1}\\
    \frac{DZ}{Dt} &= \Dot{S}_Z, \label{eq:2.4}
\end{align}
where $t$ is the time, 
$x$ is the longitudinal coordinate, 
$A$ is the cross-sectional area, 
$\rho$ is the density, 
$u$ is the velocity, 
$p$ is the pressure,
$s = \sum_{i=1}^N s_{i} Y_i$ is the non-mixing entropy, and
$D(.)/Dt = d(.)/dt + ud(.)/dx$ is the total derivative.
The right-hand side terms, ${\Dot{S_j}}$, are the sources of mass, momentum, entropy, and species, respectively. 
Because  we assume that the flow is chemically frozen with no mass generation, ${\Dot{S}_m} = 0$ and ${\Dot{S}_Z} = 0$. 
%
As shown in~\citet{jain_magri_2022}, the momentum and entropy source terms are  
\begin{align}
    &\Dot{S}_M = - \frac{4f}{{D}}\frac{u^2}{2},\\
    &\Dot{S}_s = RT\frac{f}{\zeta}\left(\frac{\gamma(1 - M^2)}{2\Lambda}\right)  \frac{DM^2}{Dt},\label{eq:EST1}
\end{align}
where {
the two/three-dimensional dissipation effects, such as recirculation and wall friction, are averaged across the cross-section and parametrized with a friction factor, $f$~\citep{jain_magri_2022},
}
$M$ is the Mach number, 
$D$ is the  diameter of the nozzle, and 
$R$ is the gas constant.  
The compressibility factor, $\Lambda$, and {\it competition factor},  $\zeta$, are defined, respectively, as
\begin{align}
&\Lambda\equiv {1 + \frac{\gamma - 1}{2}M^2},\
&\zeta \equiv  f\gamma M^2 - 2 \tan\alpha,
\end{align} 
where $\gamma$ is the heat-capacity ratio, and 
$\tan\alpha = 1/2dD/dx$ is the spatial derivative of the nozzle profile (Figure~\ref{fig:nozzle_schem}). 
%
The Gibbs equation 

\begin{align}   
T{ds} & = dh - \frac{dp}{\rho} - \sum_{i=1}^N \left(\frac{\mu_i}{W_i}\right) dY_i \label{eq:2.5}
\end{align}
closes the set of equations. The entropy of mixing is contained in the chemical potential. 
\subsection{Linearization}\label{sec:linearization}
We model the acoustics as linear perturbations that develop on top of a steady mean flow. 
For this, we decompose a generic flow variable, $v$, as $v\rightarrow {\bar v}(x) + v^{\prime}(x,t)$, where ${\bar v}(x)$ is the steady mean flow component, and $v^{\prime}(x,t)$ is the first-order perturbation. 
Linearising \eqref{eq:dmdt_1}-\eqref{eq:2.4} around the mean flow and collecting the mean flow terms yield \citep{jain_magri_2022}  

\begin{align}
    \label{eq:a2ba1m2bm1}
    \frac{A_2}{A_1} &= \frac{\bar M_1}{\bar M_2} \left(\frac{\bar \Lambda_2}{\bar \Lambda_1}\right)^\frac{(\bar \gamma + 1) \tan \alpha}{2\kappa} \left(\frac{\bar \zeta_1}{\bar \zeta_2}\right)^\frac{f\bar \gamma - 2 \tan\alpha}{2\kappa}, \\ 
       \label{eq:a2ba1m2bm13}
  \frac{\bar p_{02}}{\bar p_{01}} &= \left(\frac{\bar \Lambda_2}{\bar \Lambda_1}\right)^\frac{f\bar \gamma(\bar \gamma + 1)}{2(\bar \gamma - 1)\kappa} \left(\frac{\bar \zeta_1}{\bar \zeta_2}\right)^\frac{f\bar \gamma - 2 \tan\alpha}{2\kappa},\\
  \label{eq:delsbcpwithf}
  \frac{\Delta \bar s}{\bar c_p} & = \log\left(\frac{\bar \Lambda_1}{\bar \Lambda_2}\right)^\frac{f(\bar \gamma + 1)}{2\kappa} \left(\frac{\bar \zeta_2}{\bar \zeta_1}\right)^{\frac{\bar \gamma - 1}{\bar \gamma}\frac{f\bar \gamma - 2 \tan\alpha}{2\kappa}}, 
  %
\end{align}
where
$p_0$ is the stagnation pressure,  
$\kappa = f\bar \gamma + (\bar \gamma - 1)\tan\alpha$, and 
 $\Delta\bar s$ is the mean-flow entropy variation caused by dissipation.
The mean-flow  equations for multicomponent gases are equal to the equations of single-component gases~\citep{jain_magri_2022}. 
Physically, this is because the compositional inhomogeneities are assumed to be first-order perturbations that disturb a homogeneous mean flow. 
If the flow has no dissipation, $f=0$, the stagnation pressure and entropy in ~\eqref{eq:a2ba1m2bm13}-\eqref{eq:delsbcpwithf} are constant throughout the nozzle. 

\subsubsection{Linearisation of Gibbs equation,  density and multicomponent anisentropicity factor}
\begin{figure}
\centering    
\includegraphics[width=1\textwidth]{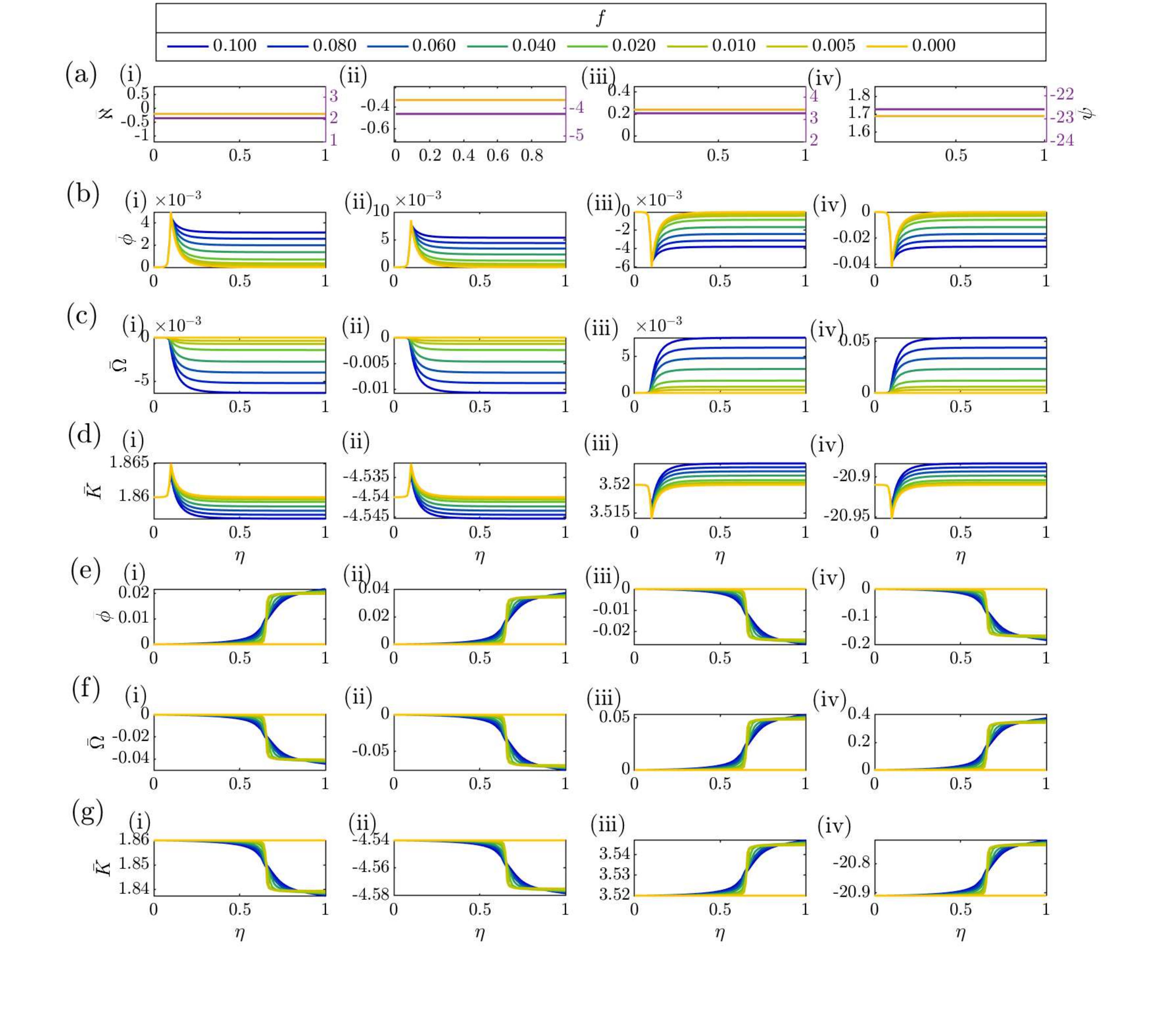}
\caption{Components of the compositional noise scaling factor, $\bar{K}$, for binary mixtures of air and (i) methane, (ii) carbon dioxide, (iii) argon,  (iv) helium. The terms in (a) do not depend on the dissipation. (b-d) subsonic flow, and (e-g) supersonic flow.
}
\label{fig:barkconst}
\end{figure}

In order to gain physical insight into the key effects that dissipation has on multicomponent gases, we linearize  and take the material derivative of the Gibbs equation \eqref{eq:2.5}

\begin{align}\label{eq:gibeqndiff}
    \frac{D}{Dt}\left(\frac{p^\prime}{\bar\gamma\bar p} - \frac{\rho^\prime}{\bar\rho} - \frac{s^\prime}{\bar c_p}\right) + \frac{c_p^\prime}{\bar c_p}\frac{D}{Dt}\left(\frac{\bar s}{\bar c_p}\right) - \frac{\gamma^\prime}{\bar\gamma}\frac{D}{Dt}\left(\frac{\bar p}{\bar\gamma \bar p}\right) - \left(\bar\aleph + \bar\psi \right) \frac{DZ^\prime}{Dt}  = 0. 
\end{align}
To track the density variation of a material fluid volume, \eqref{eq:gibeqndiff} is integrated from an unperturbed state along a characteristic line, which yields  the density
\begin{align}\label{eq:ge1}
    \frac{\rho^\prime}{\bar\rho} = \frac{p^\prime}{\bar\gamma\bar p} - \frac{s^\prime}{\bar c_p} - \bar K Z^\prime.
\end{align}
We introduce the compositional noise scaling factor, $\bar K$, as 
\begin{align}\label{eq:barKeq}
    \bar K \equiv \bar\aleph + \bar\psi + \bar \Omega + \bar\phi,
\end{align}
where $\bar\aleph$, $\bar\psi$ and $\bar\phi$ are the heat-capacity factor, chemical potential function and $\gamma^\prime$ source of noise, respectively~\citep{magri2017indirect}
\begin{align}
    \bar\psi &\equiv \frac{1}{\bar c_p \bar T} \sum_{i=1}^N \left(\frac{\bar\mu_i}{W_i} - \Delta h^o_{f,i}\right) \frac{dY_i}{dZ},\\
    \bar\aleph &\equiv \sum_{i=1}^N \left( \frac{1}{\bar{\gamma} - 1} \frac{d \log\gamma}{dY_i} + \frac{T^o}{\bar{T}} \frac{d\log c_p}{dY_i} \right) \frac{dY_i}{dZ}, \\
    \bar\phi &\equiv \sum_{i=1}^N\frac{d \log\gamma}{dY_i}{\frac{dY_i}{dZ}}\log\bar p^{\frac{1}{\bar\gamma}},\label{eq:varpi}
\end{align}
where $\Delta h^o_f$ is the standard enthalpy of formation. 
The heat capacity factor, $\bar\aleph$ and the chemical potential function, $\bar\psi$ are evaluated at the nozzle inlet~\citep{magri2017indirect}.
If the flow has a homogeneous composition, the  density depends only on fluctuations of entropy (temperature), therefore, \eqref{eq:ge1} tends to that of \citet{marble1977acoustic}.
The chemical potential function, $\bar\psi$, is typically a large contributor to compositional noise. 
Physically, when the flow is accelerated, the chemical potential energy of the compositional inhomogeneities is converted to acoustic energy \citep{magri2017indirect}.
Integrating the material derivative of the entropy fluctuation in  \eqref{eq:gibeqndiff}, considering $c_p = {\gamma R}/({\gamma - 1})$ and $\gamma^{\prime} = \sum_{i=1}^N({d\gamma}/{dY_i})(dY_i/dZ)Z^{\prime}$, yield the {\it multicomponent anisentropicity factor} 
\begin{align}
    \bar\Omega = \frac{\Delta\bar s}{\bar\gamma R}\sum_{i=1}^N\frac{d \log\gamma}{dY_i}\frac{dY_i}{dZ}
\end{align}
{where, $\Delta\bar s$ is the change in entropy of the mean flow because of dissipation.}
The dissipation considered in this work is caused by friction effects, which are encapsulated in the friction factor, $f$. Hence, by expressing the entropy variation with \eqref{eq:delsbcpwithf}, the multicomponent anisentropicity factor can be expressed as 
\begin{align}
    \bar\Omega = \frac{1}{\bar \gamma}\log\left(\frac{\bar \Lambda_1}{\bar \Lambda_2}\right)^\frac{f\bar \gamma(\bar \gamma + 1)}{2(\bar \gamma - 1)\kappa} \left(\frac{\bar \zeta_2}{\bar \zeta_1}\right)^\frac{f\bar \gamma - 2 \tan\alpha}{2\kappa}\sum_{i=1}^N\frac{d \log\gamma}{dY_i}\frac{dY_i}{dZ}. \label{eq:r438u23fkew0q}
\end{align}
The components of the compositional noise scaling factor analysed in the subsonic and supersonic regimes (\S\S\ref{sec:INtransferfunct_sub},\ref{sec:INtransferfunct_sup}) are shown in Figure \ref{fig:barkconst}.
The multicomponent anisentropicity factor is a key term that arises from the linearization of the Gibbs equation. 
It quantifies the effect that dissipation, $\Delta \bar{s}$, has on the generation of indirect noise in a multicomponent gas. 
The effect is nil when the flow is single-component, but it becomes proportionally larger as (i) the dissipation increases through $\Delta \bar{s}$, and (ii) the gas compressibility increases through $d\log\gamma/dY_i$. 
In an isentropic flow, the multicomponent anisentropicity factor is equal to zero, thus,  \eqref{eq:ge1} tends to the isentropic model of \citet{magri2017indirect}. 
To gain further physical insight into the multicomponent anisentropicity factor, we analyse the variation of the stagnation pressure 
\begin{align}\label{eq:fandpop1}
    {d\log p_0} = -\frac{\gamma {M^2}}{2} \frac{4f}{D} dx. 
\end{align}
%
For the same Mach number and friction factor, equation~\eqref{eq:fandpop1} shows that the flow inhomogeneities experience a change in the stagnation pressure depending on their composition (i.e., $\gamma = \gamma(Z)$). 
{
The difference in the stagnation pressure between the compositional inhomogeneity and the surrounding mean flow corresponds to a difference in pressure, which can be quantified by $p = {p_0}/\Lambda^\frac{\gamma}{\gamma - 1}$.  This pressure difference propagates through the nozzle as a sound wave. 
This mechanism for sound generation exists in flows with dissipation, but it does not exist in isentropic flows, in which the multicomponent anisentropicity factor is zero, i.e., $ \bar\Omega = 0$ in \eqref{eq:r438u23fkew0q}.}
This is the physical mechanism that generates indirect noise because of dissipation. 

\subsubsection{Linearized governing equations}\label{sec:goveq}

Collecting the first-order terms and using the  density \eqref{eq:ge1} yield the linearized governing equations 
\begin{align}
  &\frac{\bar D}{D\tau}\left(\frac{p^{\prime}}{\bar\gamma \bar p}\right) + \tilde{u}\frac{\partial}{\partial \eta}\left(\frac{u^{\prime}}{\bar u}\right) - \frac{\bar D}{D\tau}\left(\frac{s^{\prime}}{\bar c_p}\right)  -   \frac{D(\bar KZ^{\prime} ) }{D\tau} 
  = 0, \label{eq:masslin}\\
  &\frac{\bar D}{D\tau}\left(\frac{u^{\prime}}{\bar u}\right) + \frac{1}{\bar\gamma}\left(\frac{\tilde{u}}{\bar M^2}\right)\frac{\partial}{\partial \eta}\left(\frac{p^{\prime}}{\bar p} \right)+ \left(2\frac{u^{\prime}}{\bar u} + \frac{p^{\prime}}{\bar\gamma \bar p} (1 - \bar\gamma) - \frac{s^{\prime}}{\bar c_p} - \bar K Z^{\prime}\right)\left(\frac{4f}{\tilde{D}}\frac{\tilde{u}}{2} + \frac{\partial\tilde{u}}{\partial \eta}\right) = 0, \label{eq:momlin}\\
  & \frac{\bar D}{D\tau}\left(\frac{s^\prime}{\bar c_p}\right) = g(f,f^2,f^3, {o}(f^3)),\label{eq:entlin} \\
  &\frac{\bar DZ^\prime}{D\tau} = 0.\label{eq:specieslin}
\end{align}
 
%

The variables are non dimensionalised as $\eta = x/L$, $\tau = t f_a$, {$\tilde{D} = D/L$} and $\tilde{u} = \bar u/ c_{ref}$, where $L$ is the nozzle axial length, $f_a$ is the frequency of the advected perturbations {of the flow inhomogeneities} entering the nozzle, and $c_{ref}$ is the reference speed of sound. The non-dimensional material derivative is  $\bar D/D\tau = He \partial/\partial t + \tilde{u}\partial/\partial \eta$, where $He = {f_a L}/{c_{ref}}$ is the Helmholtz number.  
The Helmholtz number is the ratio between the wavelengths of the advected perturbations and the acoustic waves. The nozzle is compact if the wavelength of the perturbations is assumed to be infinitely larger than the length of the nozzle, i.e., $He = 0$.
The momentum equation \eqref{eq:momlin} indicates that the interaction of the inhomogeneities with the nozzle geometry and friction gives rise to noise. The dissipation term {$4f\tilde{u}/(2\tilde{D})$} in \eqref{eq:momlin} augments the effect of the acceleration of the flow.
A similar effect can be seen in \eqref{eq:entlin}. 
For an isentropic flow, the equations tend to that of \citet{magri2017indirect} in the limit of zero friction, $f \to 0$. The right hand side of equation \eqref{eq:entlin} is shown in Appendix \ref{app:fihjrf329f3}.\\

As a solution strategy, first, the primitive variables are decomposed in travelling waves  (Figure \ref{fig:nozzle_schem})
\begin{align}\label{eq:frjfr3ij}
    \pi^\pm = \frac{1}{2}\left(\frac{p^\prime}{\bar\gamma \bar p} \pm \bar M\frac{u^\prime}{\bar u} \right), \quad\quad
    \sigma = \frac{s^\prime}{\bar{c}_p}, \quad\quad
    \xi = Z^\prime.
\end{align}
Second, the partial differential equations \eqref{eq:masslin}-\eqref{eq:specieslin} are converted into ordinary differential equations by Fourier  decomposition {$(\cdot)(x,t)\to (\cdot)(x) \exp(2\pi\mathrm{i} \tau)$}, where $(\cdot)$ denotes a generic variable. (For brevity, we do not use a different notation for Fourier-transformed variables. All the variables from here on are to be interpreted as Fourier-transformed.) {Third}, the equations are cast in compact form as a linear differential equation with spatially-varying coefficients
\begin{align}
    \frac{d\mathbf{r}}{d\eta} = \left(2\pi\mathrm{i} He\mathbf{F} + \mathbf{G}\right)\mathbf{r}, \label{eq:r39234134rx}
\end{align}
where $\textbf{r} = [\pi^+, \pi^-, \sigma, \xi]^T$ is the state vector that contains the travelling waves. The matrices $\mathbf{F}$ and $\mathbf{G}$ are reported in Appendix \ref{app:matrix}. 
{Finally, \eqref{eq:r39234134rx} is solved as a boundary value problem with the bvp4c solver of MATLAB \citep{shampine2000solving}, as detailed in~\citet{jain_magri_2022}. The boundary conditions are specified according to the transfer function that is being calculated. 
The gradient, $d\mathbf{r}/d\eta$ is calculated at each spatial location from \eqref{eq:r39234134rx}, which is used to update the value of $\mathbf{r}$.
The process is repeated until the boundary conditions are satisfied (Figure~\ref{fig:nozzle_schem}).
}
Direct noise, is quantitatively measured by the acoustic-acoustic reflection and transmission coefficients, respectively 
   \begin{align}
    &R = \pi_1^- / \pi_1^+, \
    &T =\pi_2^+ / \pi_1^+. 
\end{align}
Indirect noise is quantitatively measured by the entropic-acoustic reflection and transmission, respectively 
\begin{align}
    &S_R = {\pi_1^-}/{\sigma{_1}},  \
    &S_T ={\pi_2^+}/{\sigma{_1}}, 
\end{align}
and the compositional-acoustic reflection and transmission coefficients, respectively 
\begin{align}
    &R_\xi={\pi_1^-}/{\xi{_1}},\ 
    &T_\xi ={\pi_2^+}/{\xi{_1}}. \label{eq:frjfr3ij4}
\end{align}
{Numerically, in the computation of the the entropic-acoustic transfer functions, the entropy input is assumed to be unity, $\sigma_1 = 1$, whereas the compositional input is assumed to be zero, and $\xi_1 = 0$, at the inlet. On the other hand, $\xi_1 = 1$ and $\sigma_1 = 0$ in the computation of the compositional-acoustic transfer functions. In both cases, both the right propagating acoustic wave at the inlet and the left propagating wave at the outlet are zero, $\pi_1^+ = 0$ and $\pi_2^- = 0$, respectively (Figure \ref{fig:nozzle_schem}).}
Because of the Fourier transform, the quantities in \eqref{eq:frjfr3ij}-\eqref{eq:frjfr3ij4} are complex (they have a magnitude and a phase) unless the nozzle is compact ($He=0$). 
In this work, we investigate the transfer functions for binary mixtures of four gases with air, i.e., carbon dioxide, methane, argon, and helium for flows in the subsonic regime (\S\ref{sec:INtransferfunct_sub}) and supersonic regime (\S\ref{sec:INtransferfunct_sup}). 
Table \ref{tab:cases} summarizes the cases that are being investigated.

\subsubsection{A note of caution on terminology}\label{sec:caution}
The advected wave $\sigma = s^\prime/\bar c_p$ is referred to as the entropy wave. At the nozzle inlet, by neglecting the pressure fluctuation~\citep[Eq.~(3) in][]{morgans2016entropy}, this is prescribed  as a function of the temperature fluctuation only
\begin{align}\label{eq:fe32f93d}
    \frac{s'}{\bar{c}_p} = \frac{T'}{\bar{T}}\;\;\;\;\;\;\textrm{at}\;\;\;\eta=0.
\end{align} 
In a nozzle with no dissipation, the entropy wave is only transported by the mean flow, i.e., it does not change. However, in a nozzle with friction, the entropy wave does change according to \eqref{eq:entlin}. 
To be consistent with the terminology of the literature~\citep[e.g.,][]{marble1977acoustic,morgans2016entropy}, {we refer to entropy-noise as the sound produced by the acceleration of a temperature inhomogeneity that enters the nozzle. Correspondingly, the entropic-acoustic transfer function measures the acoustic pressure that is generated by a  nozzle due to a temperature fluctuation that enters the nozzle inlet. 
} 
%

%
{\begin{table}
  \begin{center}
\def~{\hphantom{0}}
  \begin{tabular}{l|l|l|c|c}
  \multicolumn{1}{c}{Regime} &\multicolumn{1}{c}{Transfer function} &\multicolumn{1}{c}{Parameters} 
  &\multicolumn{1}{c}{Key figure} &\multicolumn{1}{c}{\S}\\
  \hline\hline
         &Compositional-acoustic ($R_\xi, T_\xi$)   &$M_t, f, \bar K$  &\ref{fig:fig13eqiv}\\[3pt]
        Subsonic &Entropic-acoustic ($S_R, S_T$)   &$He, f$  &\ref{fig:hexaxissrstentropynoise} &\ref{sec:INtransferfunct_sub}\\
         &Compositional-acoustic ($R_\xi, T_\xi$)   &$He, f, \bar K$  &\ref{fig:hexaxissrst}\\
        \hline
      Supersonic (without shock) &Compositional-acoustic ($R_\xi, T_\xi$)   &$He, f, \bar K$  &\ref{fig:choked_1} &\ref{sec:INtransferfunct_sup}\\
      Supersonic (with shock) &Compositional-acoustic   &$He, f, \bar K$  &\ref{fig:choked_2}\\
 \end{tabular}
  \caption{Summary of the cases under investigation.}
  \label{tab:cases}
  \end{center}
\end{table}}

\section{Indirect noise in subsonic nozzle flows}\label{sec:INtransferfunct_sub}
The effect of non-isentropicity is investigated for a linear-geometry nozzle in a subsonic flow. First, the effect of the friction factor is shown for a nearly compact nozzle for different throat Mach numbers. Second, the effects of the nozzle geometry and the non-compact assumption are analysed.
The equations are solved numerically with an exit temperature of $293.15 ~K$, exit pressure of $10^5 ~Pa$ and a heat-capacity ratio $\bar\gamma=1.4$. The results are calculated for a converging-diverging nozzle with the dimensions used in the experiments of \citet{de2021compositional}, which has a vena contracta factor $\Gamma = 0.89$. The nozzle has the inlet and outlet diameters of $46.2~$mm, a throat diameter of $6.6~$mm with the lengths of the converging part and diverging parts of $24~$mm and $230~$mm, respectively. The Helmholtz number is $He = 0.0074$, which is used to analyse the behaviour of the non-compact nozzle.  

\subsection{Effect of dissipation} 
\begin{figure}
\centering    
\includegraphics[width=1\textwidth]{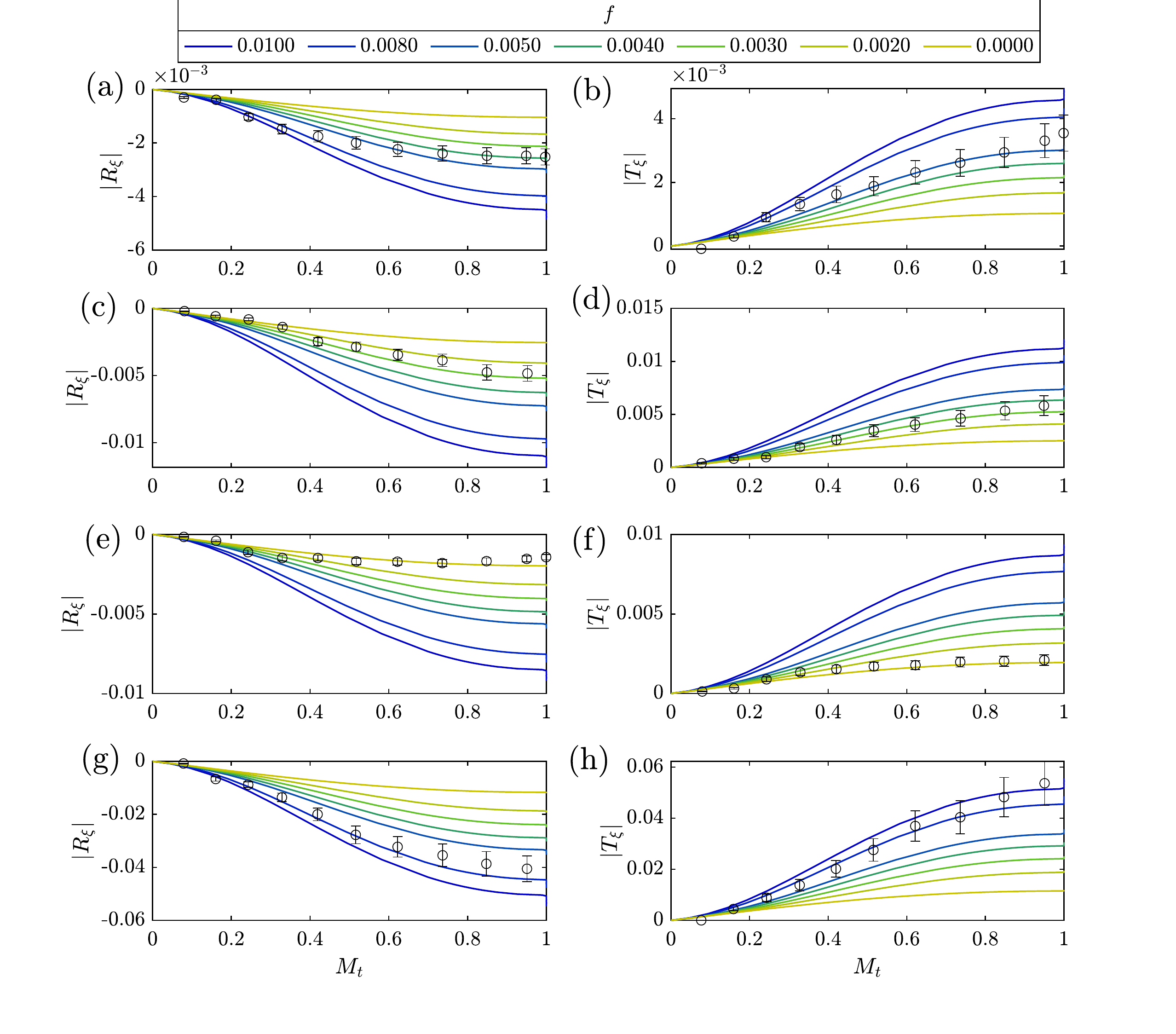}
\caption{Entropic-acoustic (left) reflection and (right) transmission coefficients (gain) for mixture of air and (a,b) carbon dioxide, (c,d) methane, (e,f) argon, (g,h) helium as a function of the throat Mach number in a nearly compact nozzle {($He = 0.0037$)}. The circles represent the experimental values 
of ~\citet{de2021compositional}.}
\label{fig:fig13eqiv}
\end{figure}
Figure \ref{fig:fig13eqiv} shows the compositional-acoustic reflection and transmission coefficients as functions of the throat Mach number and the friction factor. 
The magnitude of the reflection and transmission coefficients increases as the flow becomes more dissipative ($f$ increases). 
{
This is because  the pressure ratio across the nozzle increases with dissipation to compensate for the loss in the stagnation pressure \citep{jain_magri_2022}. This change in the pressure ratio adds to the generation of sound, which is why the magnitudes of the transfer functions tend to increase with the friction factor and the throat Mach number (e.g., Figure \ref{fig:fig13eqiv}). 
}
The trend is shared by the different gases. 
The predictions from the multicomponent model compare favourably with the experiments, the data of which is indicated by open circles with error bars. 
The predictions match different values of friction factors for different gas mixtures (Figure \ref{fig:fig13eqiv}). 
This is because, as discussed in \S\ref{sec:mathmodel}, the friction factor encapsulates the effect of dissipation in a cross-averaged sense. The amount of dissipation depends on factors such as gas composition, mass fractions, pressure and temperature. The mass fraction of the inhomogeneities is different for all gas mixtures in the experiment.

The model predictions on the reflection coefficients of mixtures of carbon dioxide and argon with air (Figures \ref{fig:fig13eqiv}a,e slightly deviate for smaller throat Mach numbers (up to $M_t \approx 0.5$). (This regime is well below the realistic Mach regime of nozzle guide vanes of aeronautical combustors~\citet{giusti2019flow}.) There are three reasons for these higher-order effects to appear.
First, friction is assumed to be constant throughout the nozzle, however, the dissipation can be different in different sections of the nozzle. A  study on the effect of a non-constant friction profile is shown in Appendix \ref{sec:appendix_frictprofile}, which shows that the friction profile can have a slight effect on the magnitude of the transfer functions. 
Second, the mass fraction of gases in the experiments ($Y_{CO_2} = Y_{Ar} = 0.2$ as compared with $Y_{CH_4} = 0.1$, and $Y_{He} = 0.02$)  may be large enough to add weakly nonlinear effects {in the acoustic propagation}. 
Third, {the effect of species diffusion and entropy/compositional dispersion \citep[e.g.,][]{mahmoudi2018low, rodrigues2020numerical} are neglected in this model, which may affect the transfer functions}. The modelling of these higher-order effects is left for future work.

\subsection{Effect of Helmholtz number}\label{sec:noncompact}
%
\begin{figure}
\centering    
\includegraphics[width=1\textwidth]{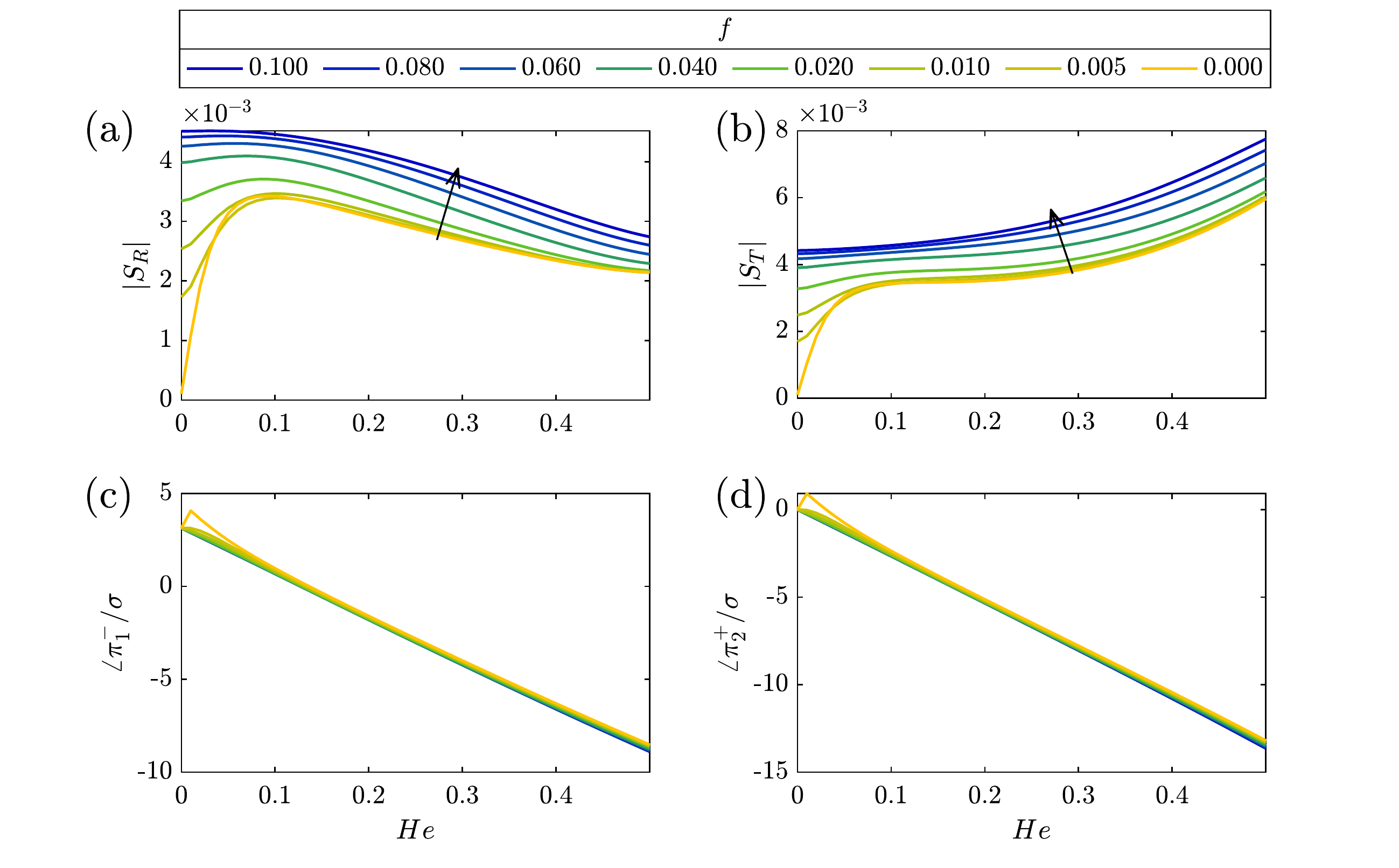}
\caption{Entropic-acoustic (a) reflection coefficient, (b) transmission coefficient, (c) phase of the reflected acoustic wave, (d) phase of the transmitted acoustic wave as a function of the Helmholtz number in a subsonic nozzle flow with throat Mach number $M_t = 0.6$.}
\label{fig:hexaxissrstentropynoise}
\end{figure}
\begin{figure}
\centering    
\includegraphics[width=1\textwidth]{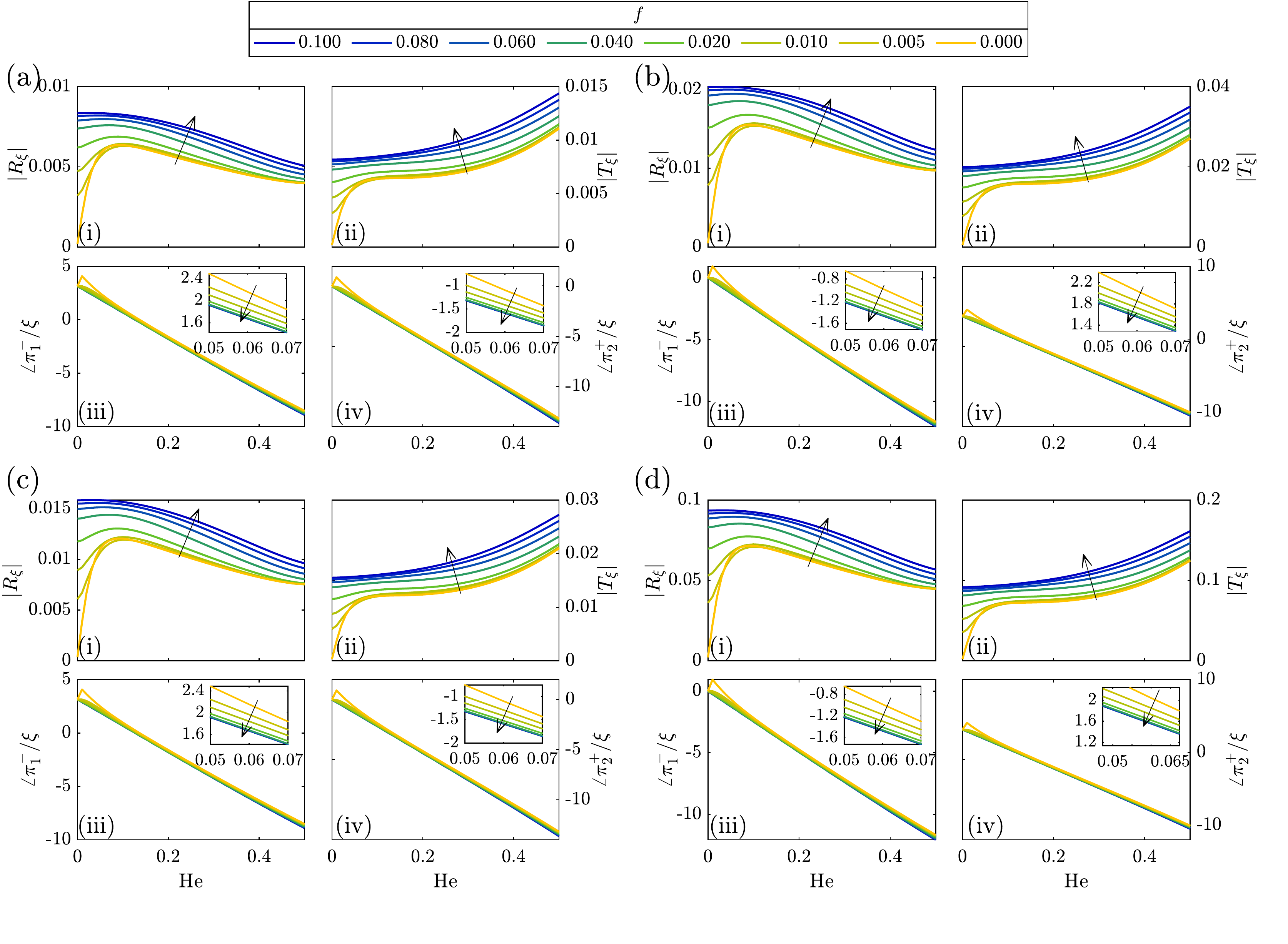}
\caption{Compositional-acoustic (i) reflection coefficient, (ii) transmission coefficient, (iii) phase of the reflected acoustic wave, (iv) phase of the transmitted acoustic wave  for mixture of air and (a) carbon dioxide, (b) methane, (c) argon and (d) helium as a function of the Helmholtz number in a subsonic nozzle flow with throat Mach number $M_t = 0.6$.}
\label{fig:hexaxissrst}
\end{figure}
%
Figure \ref{fig:hexaxissrstentropynoise} shows the gain and phase of entropic-acoustic reflection and transmission coefficients. 
As friction increases, the magnitude of the transfer function increases. 
Figure \ref{fig:hexaxissrst} shows the effect of the Helmholtz number on the gain and phase of compositional-acoustic reflection and transmission coefficients for a binary mixture of air with the four gases under consideration. 
Friction increases the magnitude of the transfer functions, but it decreases the phase. 
Different gas mixtures have different mean-flow properties that affect the amplitude, but the trends remain qualitatively similar. 

As discussed in \S\ref{sec:goveq}, the indirect noise is generated by the interaction of the flow inhomogeneities with the velocity gradient, $\partial \tilde{u}/\partial \eta$ and friction factor, $f$.
First, in an isentropic compact nozzle, the sound wave generated in the converging section is cancelled out by the wave generated in the diverging section.
 In an isentropic non-compact nozzle flow, however, the entropy and the acoustic waves have different propagation speeds, which causes a phase shift between the two. This means that the sound waves do not cancel out, hence, the acoustic coefficients are no longer equal to zero (Figures \ref{fig:hexaxissrstentropynoise} and \ref{fig:hexaxissrst}). 
Second, 
friction has a different effect on the inhomogeneities depending on whether they are in the converging or diverging section. This means that the acoustic waves do not cancel out even in a compact nozzle ($He = 0$), and larger friction leads to a larger magnitude of the transfer functions. 
Third, there is a relatively large difference in the magnitudes of the transfer functions for small Helmholtz numbers (up to $\approx 0.1$). This is because the friction induces a significant difference in phase of both reflected and transmitted waves (Figures \ref{fig:hexaxissrstentropynoise}c,d and \ref{fig:hexaxissrst}iii,iv) as compared to compact nozzle flows. This shows the importance of the non-compact assumption, in particular for small Helmholtz numbers (\S\ref{sec:sens}). 
Fourth, the transfer functions are not monotonic functions of the Helmholtz number. 
Fifth, the transfer functions are sensitive to the Helmholtz number in the vicinity of the compact nozzle, $He = 0$, for an isentropic, $f=0$, flow. This sensitivity decreases as the non-isentropicity becomes larger. This is discussed in detail in \S\ref{sec:sens}.
Finally, a phase difference of $\pi$ can be observed in the phases of the reflected and transmitted waves for carbon dioxide (Figures \ref{fig:hexaxissrst}a,iii-iv) or argon (Figures \ref{fig:hexaxissrst}b,iii-iv) and methane (Figures \ref{fig:hexaxissrst}c,iii-iv) or helium (Figures \ref{fig:hexaxissrst}d iii-iv).
This can be physically explained by analysing the chemical potential, which is the partial derivative of the Gibbs energy with respect to the number of moles of the $i$th species at constant temperature and pressure, $\mu_i = (\partial G/\partial n_i)_{p, T,n_{j\neq i}}$. The chemical potential determines the direction in which species tend to migrate \citep{job2006chemical}. 
In a mean flow of air, the chemical potential function, $\bar\psi$, is negative for carbon dioxide and argon, whereas it is positive for methane and helium. This opposite sign means that these pairs of gases have opposite tendencies to mix.
Therefore, the phases of the reflected and transmitted waves of the two pairs of gases are in antiphase (\S\ref{sec:barK}).


\section{Indirect noise in supersonic nozzles}\label{sec:INtransferfunct_sup}
\begin{figure}
\centering    
\includegraphics[width=1\textwidth]{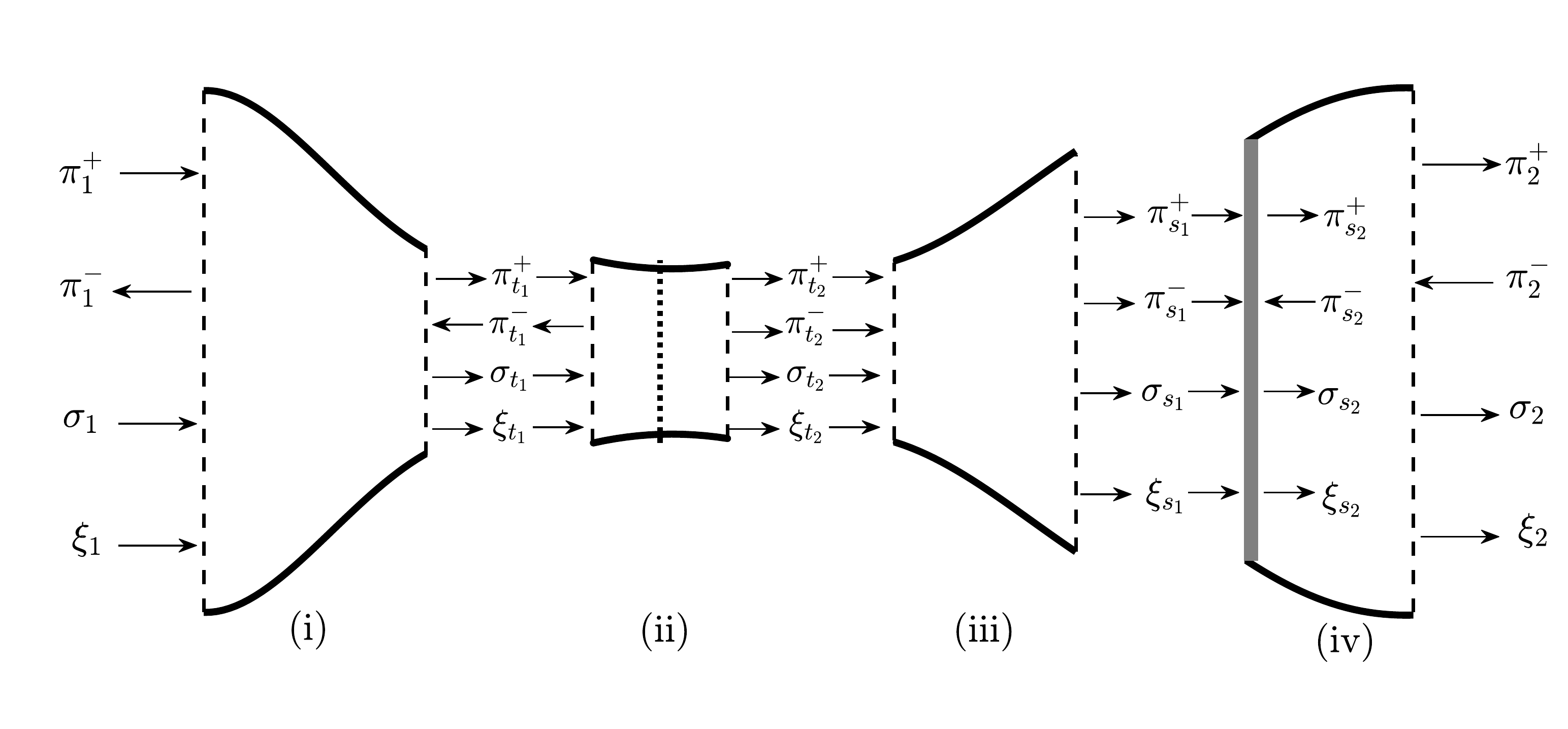}
\caption{Nozzle schematic with wave nomenclature for the supersonic regime. (i) Converging section, (ii) throat, (iii) diverging section, (iv) diverging section with a shock wave.}
\label{fig:nozz_schem_sec}
\end{figure}
We consider supersonic nozzle with a linear mean-flow velocity  with inlet and outlet Mach numbers of $0.29$ and $1.5$, respectively \citep{magri2017indirect}. The analysis is performed on  the four gas mixtures in flows without and with a shock wave. 

\subsection{Flow without a shock wave}\label{sec:chokedwoshock}

\begin{figure}
\centering    
\includegraphics[width=1\textwidth]{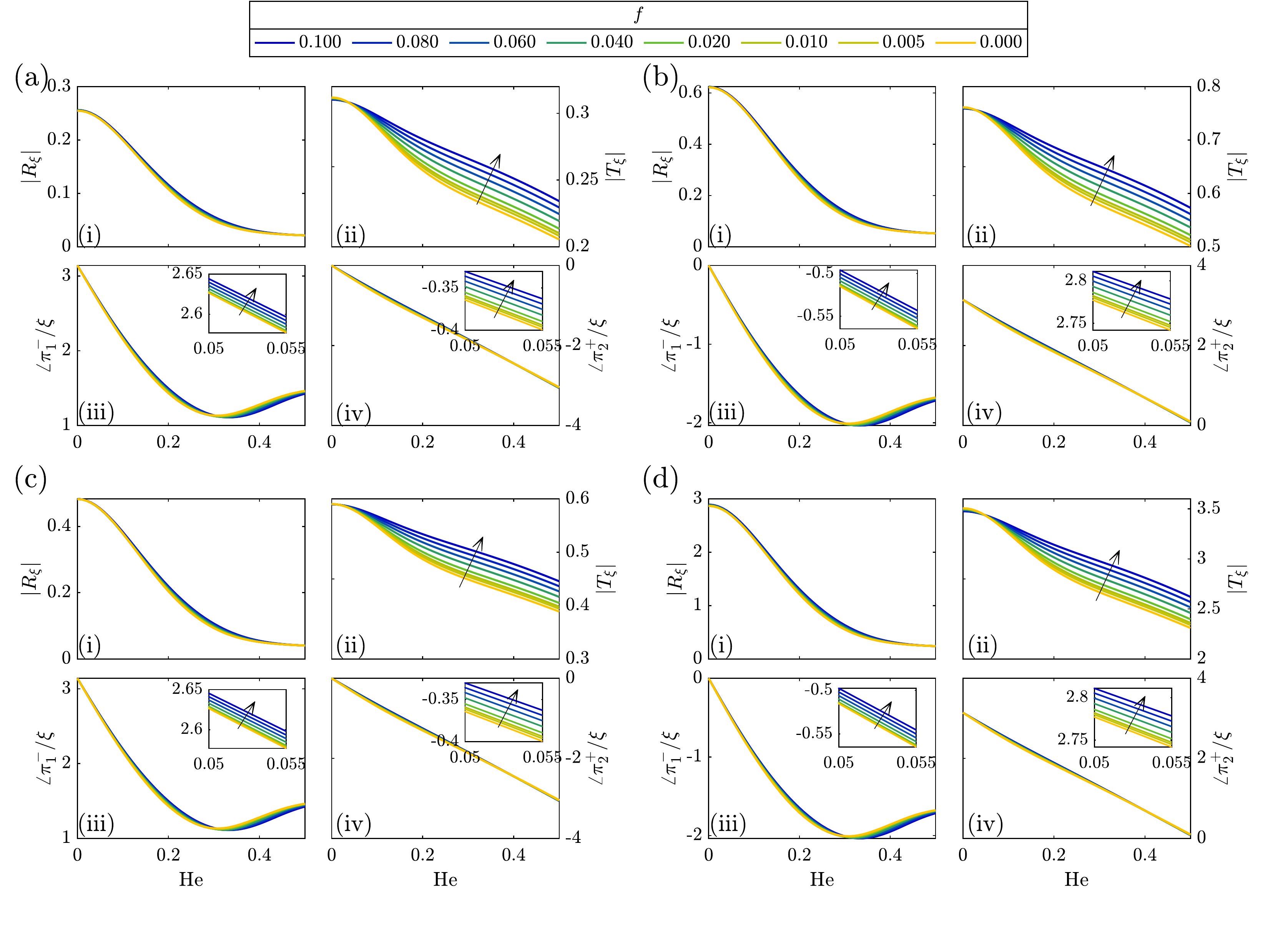}
\caption{Compositional-acoustic (i) reflection coefficient, (ii) transmission coefficient, (iii) phase of the reflected acoustic wave, (iv) phase of the transmitted acoustic wave  for mixture of air and (a) carbon dioxide, (b) methane, (c) argon and (d) helium as a function of the Helmholtz number in a supersonic nozzle flow without a shock wave.}
\label{fig:choked_1}
\end{figure}
%
%
In a choked nozzle, the upstream acoustic wave switches direction at the throat, which gives rise to a singularity in the equations. 
To deal with the singularity, the nozzle is divided into the converging section, the throat, and the diverging section, as shown in Figure~\ref{fig:nozz_schem_sec} \citep{duran2013solution}. The choking boundary condition, $M^\prime/\bar M = 0$, is imposed at the nozzle throat \citep{ magri2016compositional}   
\begin{align}
    2\frac{u^\prime}{\bar u} + \frac{p^\prime}{\bar\gamma\bar p} (1 - \bar\gamma) - \frac{s^\prime}{\bar c_p}- \bar K Z^\prime = 0.
\end{align}
The choking condition, which is affected by friction through the mean flow quantities, implicitly provides the boundary condition for the flow in the diverging section.\\

Figure \ref{fig:choked_1} shows the compositional-acoustic reflection and transmission transfer functions as functions of the Helmholtz number for different friction factors. Because the nozzle is choked, the reflected wave is largely unaffected by the friction. The magnitude of the transmission coefficient increases with the friction for higher Helmholtz numbers. It remains nearly unaffected for small Helmholtz numbers. However, the magnitude of $T_\xi$ decreases as the friction factor increases up to $He \approx 0.04$. The effect of the friction on the phase (Figures~\ref{fig:choked_1}iii,iv, inset) is opposite to the case of subsonic flow (Figures~\ref{fig:hexaxissrst}iii,iv, inset), i.e., the phase increases as the friction increases, up to $He \approx 0.3$. However, the trend reverses thereafter. A phase difference of $\pi$ is observed in the phases of carbon dioxide, methane and argon, helium mixtures with air (\S\ref{sec:noncompact}). Physically, in a subsonic flow, the pressure gradient is opposite to the flow, which makes it susceptible to flow separation. In a supersonic flow, however, the pressure gradient supports the direction of the flow both in the converging and diverging sections, which tends to alleviate dissipation effects. Therefore, the non-isentropicity on the reflected wave (for all Helmholtz numbers) and transmitted wave (for small Helmholtz numbers) is negligible in a supersonic flow without a shock wave. The friction increases the magnitude and affects the phase of the transmitted wave for larger Helmholtz numbers.

\subsection{Flow with a shock wave}
\begin{figure}
\centering    
\includegraphics[width=1\textwidth]{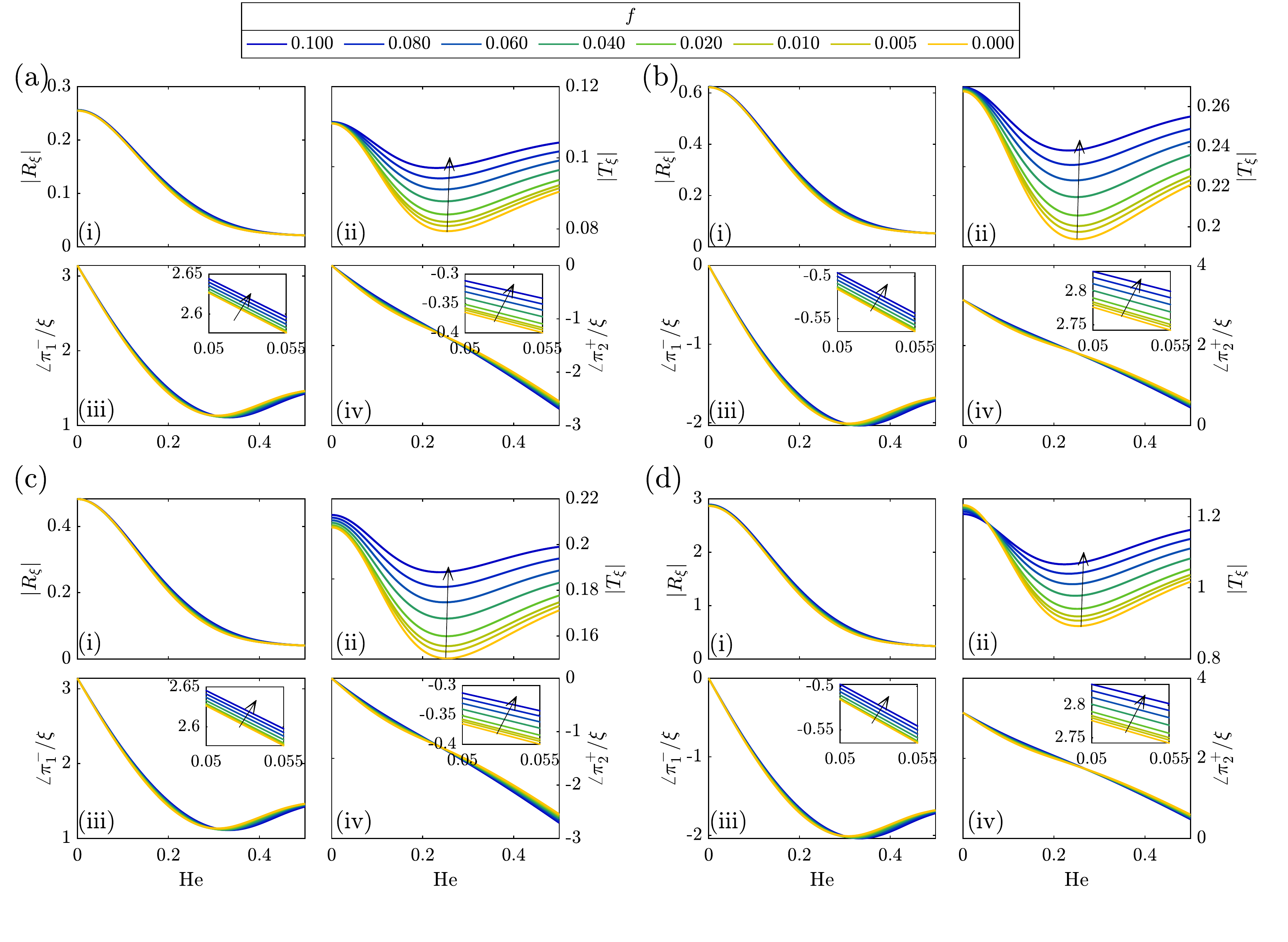}
\caption{Compositional-acoustic (i) reflection coefficient, (ii) transmission coefficient, (iii) phase of the reflected acoustic wave, (iv) phase of the transmitted acoustic wave  for mixture of air and (a) carbon dioxide, (b) methane, (c) argon and (d) helium as a function of the Helmholtz number in a supersonic nozzle flow with a shock wave.}
\label{fig:choked_2}
\end{figure}
%
%
We assume that a shock wave takes place in the diverging section (Figure \ref{fig:nozz_schem_sec}). 
The shock wave is assumed to oscillate around its mean position with an infinitesimal amplitude. 
The flow upstream of the shock is solved as described in \S\ref{sec:chokedwoshock}, and the downstream of the shock wave is solved as  described in \S\ref{sec:noncompact}. The jump conditions across the shock are imposed through the linearized Rankine-Hugoniot relations for compositional flows  \citep{magri2016compositional} 
\begin{align} \label{eq:sfhuwhrfi}
    \bar M_{s_2}^2 &= \frac{1 + \frac{\bar \gamma - 1}{2} \bar M_{s_1}^2}{\bar \gamma \bar M_{s_1}^2 - \frac{\bar \gamma - 1}{2}}, \nonumber \\ 
    \pi_{s_2}^+ &= \frac{1 + \bar M_{s_2}^2 \bar M_{s_1} + \bar M_{s_1}^2}{1 + \bar M_{s_1}^2 \bar M_{s_2} + \bar M_{s_1}^2} \pi_{s_1}^+ + \frac{1 - \bar M_{s_2}^2 \bar M_{s_1} + \bar M_{s_1}^2}{1 + \bar M_{s_1}^2 \bar M_{s_2} + \bar M_{s_1}^2} \pi_{s_1}^-, \nonumber \\ 
    \sigma_{s_2} &= \sigma_{s_1} -\left(\bar \psi_{s_2} - \bar \psi_{s_1}\right)Z^\prime+ \left( \frac{(\bar \gamma - 1)(\bar M_{s_1} - 1)^2}{\bar M_{s_1}^2(2 + (\bar \gamma - 1)\bar M_{s_1}^2)}\right)(\pi_{s_2}^+ + \pi_{s_2}^- - \pi_{s_1}^+ - \pi_{s_1}^-), 
\end{align}
where the subscripts $t$ and $s$ refer to the throat and shock wave, respectively. The subscript $1$ refers to the region just before the shock wave, whereas the subscript $2$ refers to the region just after the shock wave (Figure \ref{fig:nozz_schem_sec}).
The jump conditions \eqref{eq:sfhuwhrfi} are affected by the friction factor through the mean flow quantities. The friction, however, does not affect the linearised Rankine-Hugoniot equations. 
The composition of the gas mixture is conserved across the shock wave. 
Figure \ref{fig:choked_2} shows the variation of the transfer functions with the Helmholtz number. The reflection coefficient is equal to that of a flow without a shock wave because the nozzle is choked, which means that the information cannot travel from the diverging section to the converging section. 
The nozzle response has mixed characteristics of both supersonic and subsonic regimes. The magnitude of the transmission coefficient increases with the friction as the Helmholtz number increases. 
A qualitatively similar trend is observed in the transfer functions for different gas mixtures considered in this work, which is discussed in \S\ref{sec:barK}. 

\section{Semi-analytical solution}\label{sec:asymptexpn}
We propose a semi-analytical solution to estimate the nozzle transfer functions using an asymptotic expansion  based on path integrals \citep{magri2017indirect,magri2023linear}. 
This is useful for perturbation and sensitivity analysis (\S\ref{sec:sens}). 
First, the differential equation \eqref{eq:r39234134rx} is cast in integral form as
\begin{align}\label{eq:AE_integ}
    \mathbf{r(\eta)} = \mathbf{r}_a + 2\pi i  He \int_{\eta_a}^\eta \mathbf{F(\eta^\prime)}\mathbf{r(\eta^\prime)} d\eta^\prime +  \int_{\eta_a}^\eta \mathbf{G(\eta^\prime)}\mathbf{r(\eta^\prime)} d\eta^\prime,
\end{align}
where $\eta_a$ is the nozzle axial coordinate of the inlet, $\eta_a = 0$.
In this case, the asymptotic expansion can be performed on the Riemann invariant, $\mathbf{r(\eta)}$ by using \eqref{eq:AE_integ}. 
The solution can be formally written
as
\begin{align}
    \mathbf{r(\eta)} = \mathbf{H(\eta)}\mathbf{r}(\eta_a), 
\end{align}
where $\mathbf{r}(\eta_a)$ is the solution at inlet, and $\mathbf{H}$ is the propagator. When the commutator $\mathbf{H}(\eta_1)\mathbf{H}(\eta_2) - \mathbf{H}(\eta_2)\mathbf{H}(\eta_1)$ is zero, the solution is obtained by integrating \eqref{eq:AE_integ} term by term to give an exponential. 
However, when the acoustic commutator is not zero, which is the case here, the formal solution \eqref{eq:AE_integ}  is substituted recursively in \eqref{eq:r39234134rx}, which yields an explicit solution in an asymptotic form 
\begin{align}\label{eq:AE_bigeqn}
    &\mathbf{r(\eta)} = \left[\sum_{k = 0}^{n}(2\pi i  He)^k \sum_{i = 0}^{n - k} \mathcal{I}\{ \mathbf{F}^k\mathbf{G}^i\}\right]\mathbf{r(\eta_a)},
\end{align}
where we define $\mathcal{I}$ as
\begin{align}
    \mathcal{I}\{ \mathbf{F}^k\mathbf{G}^i\} = \int_{\eta_a}^{\eta}d\eta^{(1)}\ldots\int_{\eta_a}^{\eta^{(i+k)}}d\eta^{(i+k-1)} \mathcal{P}\{ \mathbf{F}^k\mathbf{G}^i\},
\end{align}
in which the integrals are path-ordered (Dyson-Feynman integrals), $\eta_a<\eta^{(n)}<\ldots<\eta^{(1)}<\eta$, and $\mathcal{P}\{ \mathbf{F}^k\mathbf{G}^i\}$ is the path-ordered multiplication of all combinations of the matrices.
For instance, $\mathcal{P}\{ \mathbf{F}^2\mathbf{G}^1\} = \mathbf{F}(\eta^{(1)})\mathbf{F}(\eta^{(2)})\mathbf{G}(\eta^{(3)}) + \mathbf{F}(\eta^{(1)})\mathbf{G}(\eta^{(2)})\mathbf{F}(\eta^{(3)}) + \mathbf{G}(\eta^{(1)})\mathbf{F}(\eta^{(2)})\mathbf{F}(\eta^{(3)})$. 
An explicit expression of \eqref{eq:AE_bigeqn} is shown in Appendix \ref{app:AE}.
In a finite spatial domain, the solution of \eqref{eq:AE_bigeqn} is convergent when $\mathbf{F}$ and $\mathbf{G}$ are bounded \citep{lam1998decomposition}. The boundary conditions are computed similarly to  \citet{duran2013solution}. In a supersonic flow, the equations are solved as a subsonic flow from the inlet to the nozzle throat with a choking condition at the throat.
The final solution can be obtained by computing $\mathbf{H(\eta)}$ analytically or numerically using different methods \citep{moler2003nineteen}. 
\section{Compositional noise scaling factor}\label{sec:barK}

From the analysis of \S\S\ref{sec:INtransferfunct_sub},\ref{sec:INtransferfunct_sup}, we observe qualitatively similar trends between
(i) the compositional-noise transfer functions (Figure \ref{fig:hexaxissrst}) and the entropy-noise transfer functions for single-component gases (Figure \ref{fig:hexaxissrstentropynoise}); and 
(ii) the transfer functions of different gas mixtures.
Mathematically, we observe that the source terms $s^\prime$ and $\bar{K} Z^\prime$ appear side by side in the linearised Gibbs equation \eqref{eq:ge1} and the governing equations \eqref{eq:masslin}-\eqref{eq:specieslin}. 
Because the partial differential equations~\eqref{eq:masslin}-\eqref{eq:specieslin} are linear, the transfer functions are approximately related to each other through the scaling factor $\bar{K}$ as 
\begin{align}\label{eq:scalingen}
    &T_\xi \approx \bar K S_T, \
    &R_\xi \approx \bar K S_R,  
\end{align}
which enable the estimation of the compositional-noise transfer functions from the knowledge of one reference transfer function of a single-component gas. 
In this section, we numerically show the accuracy of this scaling for a range of mixtures and friction factors. 

\begin{figure}
\centering    
\includegraphics[width=1.0\textwidth]{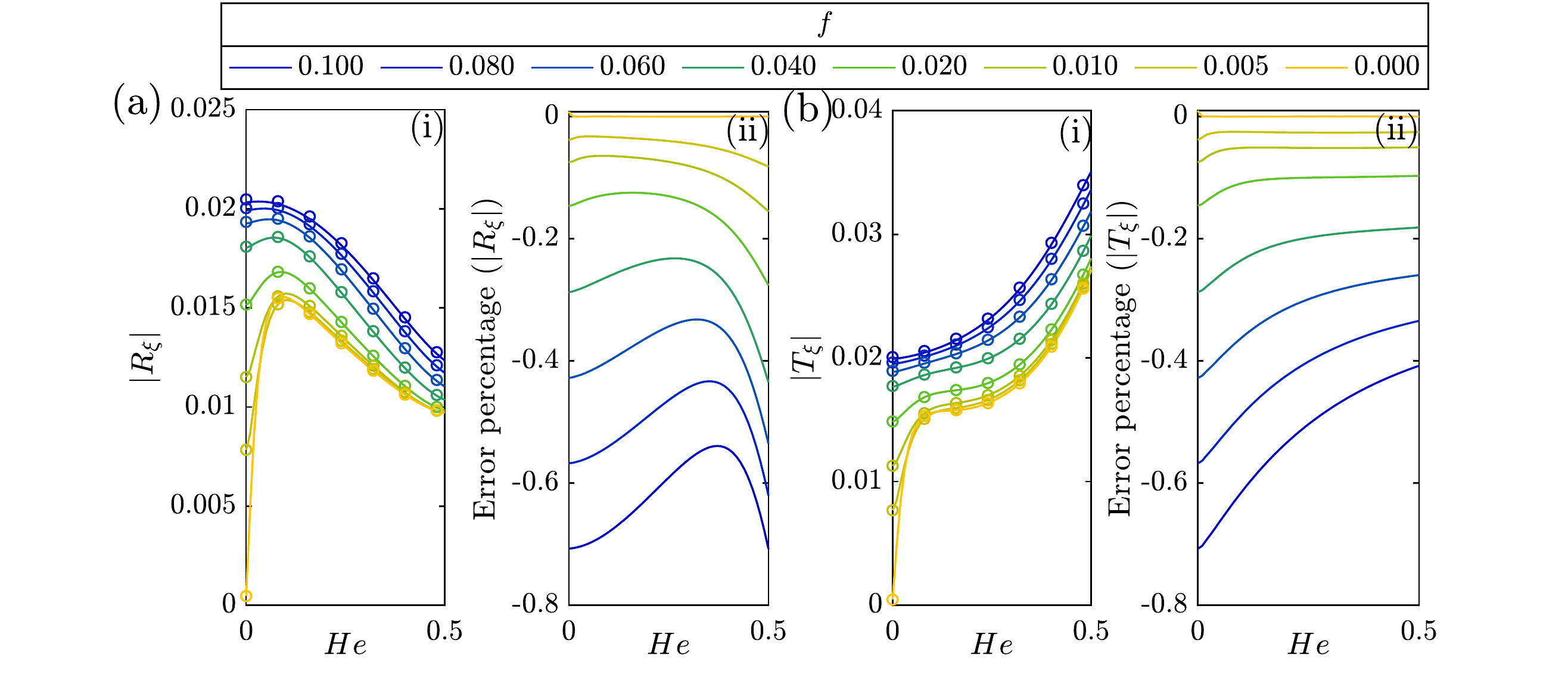}
\caption{(i) Compositional-acoustic reflection (left) and transmission (right) coefficients, (ii) error percentage for a mixture of air and methane for subsonic flow.
The circles correspond to the coefficients calculated from the single-component entropic-acoustic transfer functions scaled with $\bar K_{CH_4}$.}
\label{fig:scaling_entropynoise}
\end{figure}

 Figure \ref{fig:scaling_entropynoise}i shows the estimated values of the compositional-noise transfer functions calculated from the single-component entropic-acoustic reflection coefficient (circles) from \eqref{eq:scalingen}. The agreement is satisfactory because the absolute error is smaller than $1\%$.  
When one compositional-noise transfer function is available for mixture, say, 1, we can compute the transfer functions for another mixture, say 2, as  
\begin{align}
    &T_{\xi_{mix_1}} = \frac{\bar K_{mix_1}}{\bar K_{mix_2}} T_{\xi_{mix_2}}, \
    &R_{\xi_{mix_1}} = \frac{\bar K_{mix_1}}{\bar K_{mix_2}} R_{\xi_{mix_2}}. 
\end{align}
For illustration, we consider mixtures of methane and carbon dioxide only. 
The error between the benchmark solution, which is obtained by solving the governing equations, and the scaled solution is defined as 
\begin{align}
    \text{Error $\%$} = \frac{\left|R_{\xi_{CH_4}}\right| - \frac{\bar K_{CH_4}}{\bar K_{CO_2}} \left|R_{\xi_{CO_2}}\right|}{\left|R_{\xi_{CH_4}}\right|} \times 100.
\end{align}
\begin{figure}
\centering    
\includegraphics[width=1.0\textwidth]{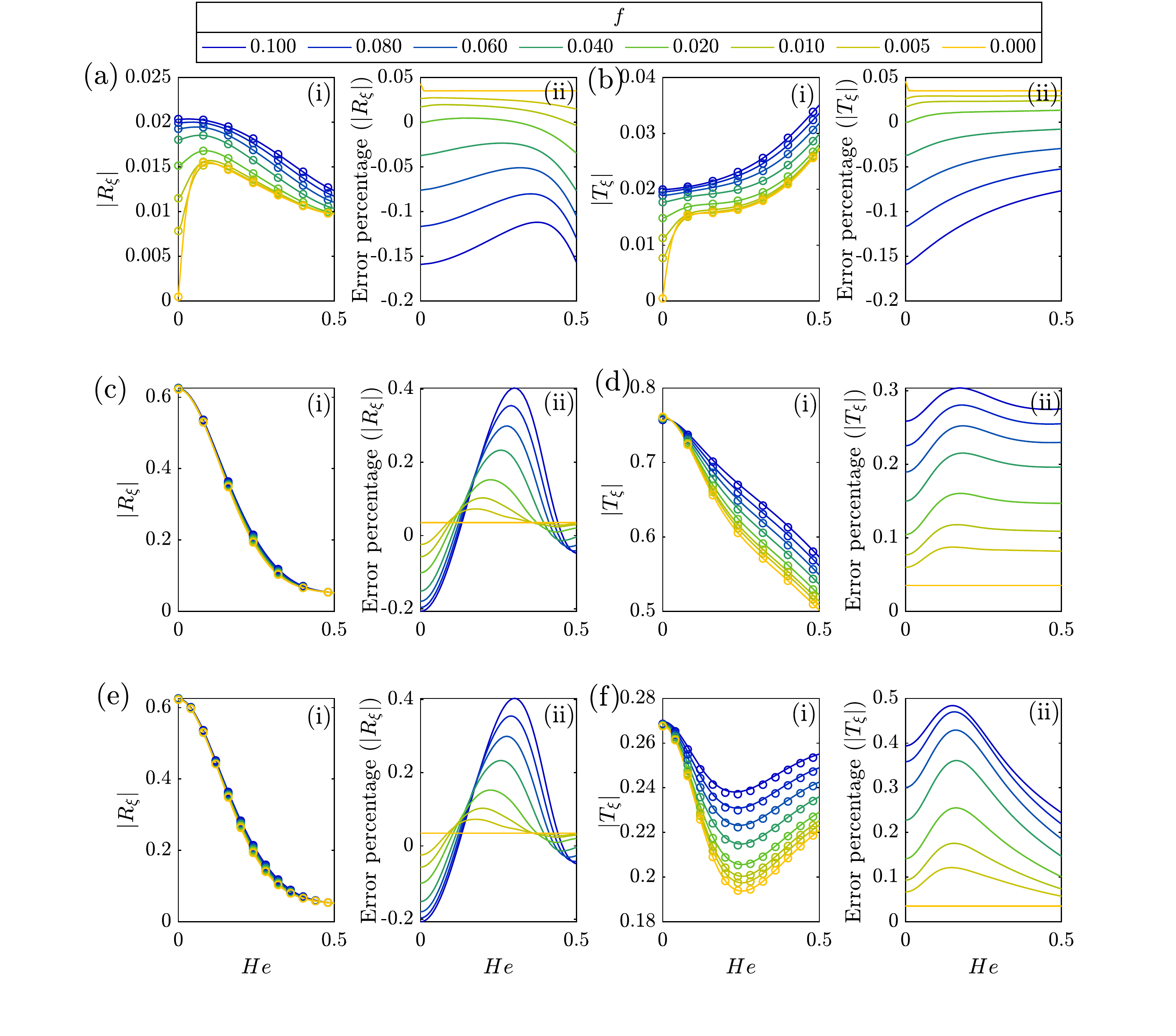}
\caption{(i) Compositional-acoustic reflection (left) and transmission (right) coefficients, (ii) error percentage for a mixture of air and methane for (a,b) subsonic, (c,d) supersonic without a shock wave and (e,f) with a shock wave.
The circles correspond to the coefficients calculated from the transfer function of $CO_2$-air mix scaled with $\bar K_{CH_4}/\bar K_{CO_2}$.}
\label{fig:choked_11}
\end{figure}
Figure \ref{fig:choked_11} shows the transfer functions for a  methane-air mixture using the carbon dioxide-air mixture and the compositional noise scaling factor (shown by circles). 
The solid lines are the results obtained by solving the model proposed in \S\ref{sec:mathmodel}. 
The comparison is performed for subsonic (Figures~\ref{fig:choked_11}a,b), supersonic without a shock wave (Figures~\ref{fig:choked_11}c,d) and supersonic with a shock wave (Figures~\ref{fig:choked_11}e,f) regimes.
The exact results closely match the scaled results for both cases. The absolute error is smaller than $0.5\%$. 
%
%
\begin{table}
  \begin{center}
\def~{\hphantom{0}}
  \begin{tabular}{l c c}
{} &Gas mixture & $\bar K$\\
  \hline
     (a) &$CO_2$   & $> 0$\\
      (b) &$CH_4$   & $< 0$\\
      (c) &$Ar$   & $> 0$\\
      (d) &$He$   & $< 0$\\
 \end{tabular}
  \caption{Compositional noise scaling factor, $\bar K$ for the gas mixtures used in the study.   
  }
  \label{tab:barK}
  \end{center}
\end{table}
Additionally, we observe a phase difference of $\pi$ in the transfer functions for different gas mixtures in Figures \ref{fig:hexaxissrst}, \ref{fig:choked_1}, and \ref{fig:choked_2}. The phase depends on the sign of the compositional noise scaling factor $\bar K$ (Table \ref{tab:barK}), as explained in~\S\ref{sec:noncompact}.  
In summary, the compositional noise factor, $\bar K$ can be used to quickly estimate the compositional-noise transfer functions from a reference transfer function. 

\section{Conclusions}
In this work, we propose a physics-based  model to calculate indirect-noise transfer functions in a flow with dissipation and multicomponent gases. The sources of viscous dissipation  averaged over the cross-section are encapsulated by a friction factor.
First, we show the effect that dissipation has on a multicomponent subsonic flow. We compute the transfer functions, which favourably compare with the experiments available in the literature. 
Second, we extend the model to supersonic flows with and without a shock wave. 
Dissipation has the effect of increasing the gain of transfer functions for higher Helmholtz numbers.
Third, we propose a semi-analytical solution to calculate the transfer functions with an asymptotic expansion. 
Fourth, we observe that the transfer functions have a similar trend for different gas mixtures. Thus, we introduce a compositional noise scaling factor, which is employed to cheaply estimate the transfer functions for any gas mixture from the knowledge of a reference transfer function.
%
The proposed low-order model can be used to estimate the nozzle transfer functions, which is useful for preliminary design. This work opens up new possibilities for accurate modelling of sound generation in aeronautics and power generation with realistic nozzles. 

\section*{Acknowledgements}
A. Jain is supported by the University of Cambridge Harding Distinguished Postgraduate Scholars Programme. L. Magri gratefully acknowledges financial support from the ERC Starting Grant PhyCo 949388.

\appendix
\section{Linearised entropy source term}\label{app:fihjrf329f3}

The right-hand-side term of equation \eqref{eq:entlin} is 
\begin{align}
&g(f,f^2,f^3, o(f^3)) = C_I\left(\frac{\bar M^2(\bar \gamma - 1)}{2}\frac{\partial}{\partial \eta}\left(\frac{u^{\prime}}{ \bar u}\right) - \frac{\partial}{\partial \eta}\frac{p^{\prime}}{\bar \gamma \bar p}\right) + C_{II} \left(2\frac{u^{\prime}}{ \bar u} + (1 - \bar \gamma) \frac{p^{\prime}}{\bar \gamma \bar p} - \frac{s^\prime}{\bar c_p} - \bar K Z^\prime\right), \\
    &C_I = \Theta f\left(\bar \gamma \bar M^2 f - 2 \tan\alpha \right),\\
    &C_{II} = \Theta \Bigg(2 \tan\alpha \bar M \frac{d \bar M}{d\eta}\left(\frac{1 - 
    (\bar \gamma + 2) \bar M^2}{(1 - \bar M^2)\Lambda} + 2\right) f  \ldots \nonumber\\
    \ldots& - \left(\bar \gamma \bar M^3 \frac{d \bar M}{d\eta}\left(\frac{1 - 
    (\bar \gamma + 2) \bar M^2}{(1 -  \bar M^2)\Lambda}\right) - \frac{4\tan\alpha}{\tilde{D}} \bar M^2 \right)f^2 - \frac{2\bar \gamma \bar M^4}{\tilde{D}}f^3\Bigg),
\end{align}
and
\begin{align}
    &\Theta = \tilde{u}\frac{2(\bar \gamma - 1) (1 -  \bar M^2)}{8\tan^2\alpha\Lambda - 2 \tan\alpha\bar \gamma \bar M^2(\bar \gamma - 1) (1 -  \bar M^2)f + (\bar \gamma + 1)\bar \gamma^2  \bar M^4 f^2}.
\end{align}


\section{Matrix formulation}\label{app:matrix}
The terms of matrices $\mathbf{F}$ and $\mathbf{G}$ are
\begin{align}
    \mathbf{F} = \begin{bmatrix}
           \varrho^+ &\varkappa^+   &0 &0\\
           \varkappa^- &\varrho^-   &0 &0\\
           \vartheta^+ &\vartheta^-   &\varsigma &0\\
           0 &0   &0 &-\frac{1}{\bar u}\\
         \end{bmatrix}, 
    \quad\quad\mathbf{G} = \begin{bmatrix}
          \Gamma^-\kappa^- + \mathcal{M} &\Gamma^-\kappa^+ - \mathcal{M}   &\Gamma^- &\bar K \Gamma^-\\
          \Gamma^+\kappa^- - \mathcal{M} &\Gamma^+\kappa^+ + \mathcal{M}   &\Gamma^+ &\bar K \Gamma^+\\
          \Upsilon\kappa^- &\Upsilon\kappa^+   &\Upsilon &\bar K \Upsilon\\
          0 &0   &0 &0\\
         \end{bmatrix}
\end{align}
where
\begin{align}
    \varrho^\pm &= \frac{\pm\bar M \left(2(1 \mp \bar M) + C_1 \bar M \left(\bar M \mp 2 + \bar M\gamma(1 \mp 2\bar M)\right) \right)}{2\left(\bar  M^2 \bar u - \bar u + C_1\bar M^2\bar u + C_1 \bar M^4 \gamma \bar u\right)}\\
    \varkappa^\pm &= \frac{\bar M^2 C_1 \left( 2 \pm {\bar M (\gamma - 1)}\right)}{2\left(\bar  M^2 \bar u - \bar u + C_1\bar M^2\bar u + C_1 \bar M^4 \gamma \bar u\right)}\\
    \vartheta^\pm &= \frac{C_1\bar M \left(\bar M(1 + \gamma) \pm \bar M^2 (1 - \gamma) \mp 2\right)}{\left(\bar  M^2 \bar u - \bar u + C_1\bar M^2\bar u + C_1 \bar M^4 \gamma \bar u\right)}\\
     \varsigma &= -\frac{C_1\bar M^2 \left(1 + \gamma\bar M^2\right) + \bar M^2 - 1}{\left(\bar  M^2 \bar u - \bar u + C_1\bar M^2\bar u + C_1 \bar M^4 \gamma \bar u\right)}\\
    \mathcal{M} &= \frac{1}{2\bar M}\frac{d\bar M}{dx}, \quad\quad
    \kappa^\pm = {(\gamma -1) \pm \frac{2}{M}}\\
    \Gamma^\pm &= \bar M\frac {\left(C_a\left(\bar M \pm 1\right) \pm F_f\bar M \left(C_1 \gamma \bar M^3 + \bar M \pm \left( 1 - C_1 \bar M^2\right)  \right)\right) }{2(\bar  M^2 \bar u - \bar u + C_1\bar M^2\bar u + C_1 \bar M^4 \gamma \bar u)}\\
    \Upsilon &= \frac{\left( C_a(\bar M^2 - 1) - C_1 F_f \bar M^4 (1 + \gamma)\right)}{(\bar  M^2 \bar u - \bar u + C_1\bar M^2\bar u + C_1 \bar M^4 \gamma \bar u)}\\ 
    C_1 &= \frac{(\bar\gamma - 1)(1 - \bar M^2) f}{2(1 + \frac{\bar\gamma - 1}{2}\bar M^2)(\bar\gamma \bar M^2 f - 2 \tan\alpha)}\\
    F_f &= -\left(\frac{\partial\bar u}{\partial x} + \frac{4f}{\tilde{D}}\frac{\bar u}{2} \right), \quad\quad 
    C_a = -C_1 \bar M \bar u \frac{d\bar M}{d x} \left(2 - \frac{2\bar M^2}{1 - \bar M^2} - \frac{\bar C}{\bar A}\right)\\
    \bar A &= 2\left(1 + \frac{\bar\gamma - 1}{2}\bar M^2\right)\left(\bar\gamma \bar M^2 f - 2 \tan\alpha\right)\\
    \bar C &= 2\bar\gamma \bar M^2\left(-2f\left(1 + \frac{\bar\gamma - 1}{2}\bar M^2\right) + \frac{\bar\gamma - 1}{\bar\gamma}\left(2 \tan\alpha - \bar\gamma \bar M^2 f\right)\right)
\end{align}
\section{Effect of the friction  profile}\label{sec:appendix_frictprofile}
\FloatBarrier
\begin{figure}
\centering    
\includegraphics[width=0.8\textwidth]{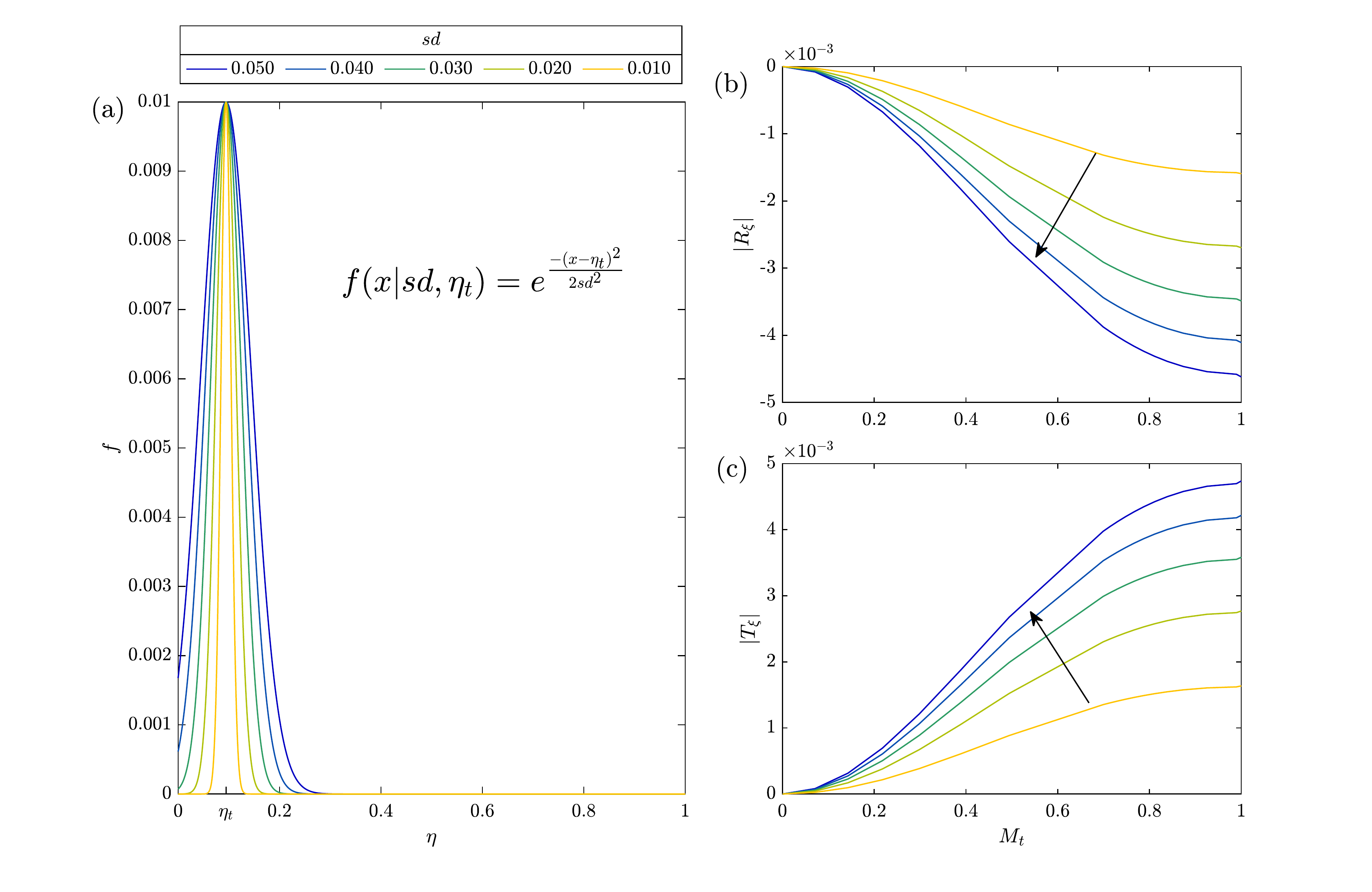}
\caption{Effect of friction profile. (a) Friction as a function of the spatial distance. (b) Reflection and (b) transmission coefficients as a function of the throat Mach number for different standard deviations, $sd$.
}
\label{fig:EoFP_1}
\end{figure}
{The friction factor is assumed to be a constant quantity in the analysis presented in this work. It takes into account sources of dissipation averaged across the nozzle cross-section. Depending on the flow conditions and the nozzle shape, the friction factor might spatially vary. In this section, we assume that the dissipation is concentrated near the throat, where most of the dissipation occurs~\citep{jain_magri_2022}. We model the friction profile as a Gaussian }
\begin{align}
    f({x|sd}, \eta_t) = e^\frac{-(x - \eta_t)^2}{2{sd}^2},
\end{align}
where $\eta_t$ is throat location.
The variable $sd$ defines the spread of the friction near the throat. Figure \ref{fig:EoFP_1}a shows the friction profile for different values of spread, $sd$. Figures  \ref{fig:EoFP_1}b,c  show the effect of the spread, $sd$, on the reflection and transmission coefficients. As the spread increases, the magnitude increases due to a net increase in the effect of friction.
For different flow conditions, the friction factor and its distribution can be approximated from the experiments. 
%
%
%
\FloatBarrier
\section{Semi-analytical solution}\label{app:AE}

The explicit expression of \eqref{eq:AE_bigeqn} is
\begin{align}\label{eq:AE_bigeqn_big}
    &\mathbf{r(\eta)} = \Bigg[ \mathrm{1} + \int_{\eta_a}^{\eta}d\eta^{(1)}\mathbf{G}(\eta^{(1)}) + \ldots 
    + \int_{\eta_a}^{\eta}d\eta^{(1)}\ldots\int_{\eta_a}^{\eta^{(n-1)}}d\eta^{(n)}\mathbf{G}(\eta^{(1)})\mathbf{G}(\eta^{(2)})\ldots\mathbf{G}(\eta^{(n)})   \nonumber\\
     &+ 2\pi\iota He \Bigg(\int_{\eta_a}^{\eta}d\eta^{(1)}\mathbf{F}(\eta_1)+  \int_{\eta_a}^{\eta}d\eta^{(1)} \int_{\eta_a}^{\eta^{(1)}}d\eta^{(2)}\bigg[\mathbf{F}(\eta^{(1)})\mathbf{G}(\eta^{(2)}) + \mathbf{G}(\eta^{(1)})\mathbf{F}(\eta^{(2)})\bigg] 
     +\ldots \nonumber\\
     &+  \int_{\eta_a}^{\eta}d\eta^{(1)}\ldots\int_{\eta_a}^{\eta^{(n-1)}}d\eta^{(n)}\bigg(\mathbf{F}(\eta^{(1)})\mathbf{G}(\eta^{(2)})\ldots\mathbf{G}(\eta^{(n)}) + \mathbf{G}(\eta^{(1)})\mathbf{F}(\eta^{(2)})\ldots\mathbf{G}(\eta^{(n)}) \nonumber\\
     &+ \ldots + 
     \mathbf{G}(\eta^{(1)})\mathbf{G}(\eta^{(2)})\ldots\mathbf{F}(\eta^{(n)})\bigg) \Bigg) +\ldots \nonumber\\
     &+ \left(2\pi\iota He\right)^{n-1} \Bigg( \int_{\eta_a}^{\eta}d\eta^{(1)}\ldots\int_{\eta_a}^{\eta^{(n-2)}}d\eta^{(n-1)}\mathbf{F}(\eta^{(1)})\mathbf{F}(\eta^{(2)})\ldots\mathbf{F}(\eta^{(n-1)}) + \ldots \nonumber\\
     &+  \int_{\eta_a}^{\eta}d\eta^{(1)}\ldots\int_{\eta_a}^{\eta^{(n-2)}}d\eta^{(n - 1)}\bigg(\mathbf{F}(\eta^{(1)})\mathbf{F}(\eta^{(2)})\ldots\mathbf{F}(\eta^{(n-2)})\mathbf{G}(\eta^{(n-1)}) +\nonumber\\
     & \mathbf{F}(\eta^{(1)})\ldots\mathbf{G}(\eta^{(n - 2)})\mathbf{F}(\eta^{(n - 1)}) + \ldots + 
     \mathbf{G}(\eta^{(1)})\mathbf{F}(\eta^{(2)})\ldots\mathbf{F}(\eta^{(n-1)})\bigg) \Bigg) \nonumber\\
     &+ \left(2\pi\iota He\right)^{n} \Bigg( \int_{\eta_a}^{\eta}d\eta^{(1)}\ldots\int_{\eta_a}^{\eta^{(n-1)}}d\eta^{(n)}\mathbf{F}(\eta^{(1)})\mathbf{F}(\eta^{(2)})\ldots\mathbf{F}(\eta^{(n)})\Bigg]\mathbf{r}(\eta_a)
\end{align}

\section{{Perturbation analysis} of the acoustic transfer functions}\label{sec:sens}
\FloatBarrier
In \S\S\ref{sec:INtransferfunct_sub},\ref{sec:INtransferfunct_sup}, we numerically evaluate the acoustic transfer functions for a range of Helmholtz numbers. 
{
In \S \ref{sec:asymptexpn}, we propose a semi-analytical solution based on path integrals as a semi-analytical solution approach by series expansion. 
Here, we investigate the leading perturbation effects that changes in the Helmholtz number have on the transfer functions.
We show the results for the compositional-acoustic reflection coefficients---similar  conclusions can be drawn for the transmission coefficient. 
}
{Perturbation analysis is performed by evaluating each order of the asymptotic solution~\eqref{eq:AE_bigeqn}, each term of which is equal to the corresponding order of the Taylor expansion at the expansion point $He_0$~\citep{magri2023linear}. 
}   

\begin{figure}
\centering    
\includegraphics[width=0.8\textwidth]{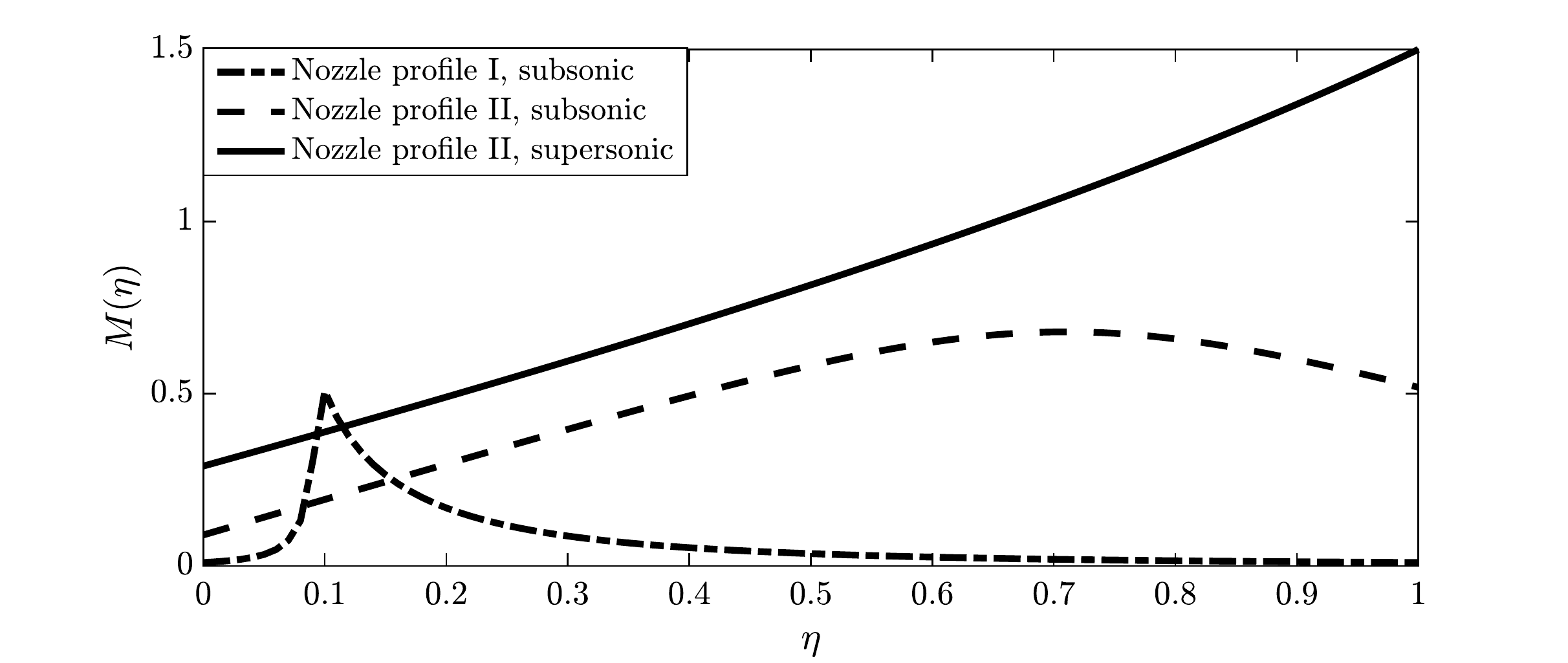}
\caption{{Mach number as a function of the  nozzle coordinate, $\eta$, for three cases; (i) Nozzle profile I, subsonic flow in a linear geometry nozzle, $M_1 = M_2 = 0.01$ and $M_t = 0.6$;
(ii) Nozzle profile II, subsonic flow, $M_1 = 0.09$, $M_2 = 0.5$, and $M_t = 0.7$;
(iii) Nozzle profile II, supersonic flow,  $M_1 = 0.29$, $M_2 = 1.5$ without dissipation. 
}}
\label{fig:machnoprofiles_taylor}
\end{figure}

{
We analyse the effect of  perturbations to the Helmholtz number for three flows as shown in Figure \ref{fig:machnoprofiles_taylor}. 
First, a subsonic flow in a nozzle profile is investigated in \S \ref{sec:INtransferfunct_sub} with equal inlet and outlet areas (Nozzle profile I). 
Second, we investigate a subsonic flow in a nozzle with the profile of \S\ref{sec:INtransferfunct_sup} (Nozzle profile II) with an outlet area that is half of the inlet area. 
Third, we analyse the Nozzle profile II for a supersonic regime.
}
\begin{figure}
\centering    
\includegraphics[width=1.0\textwidth]{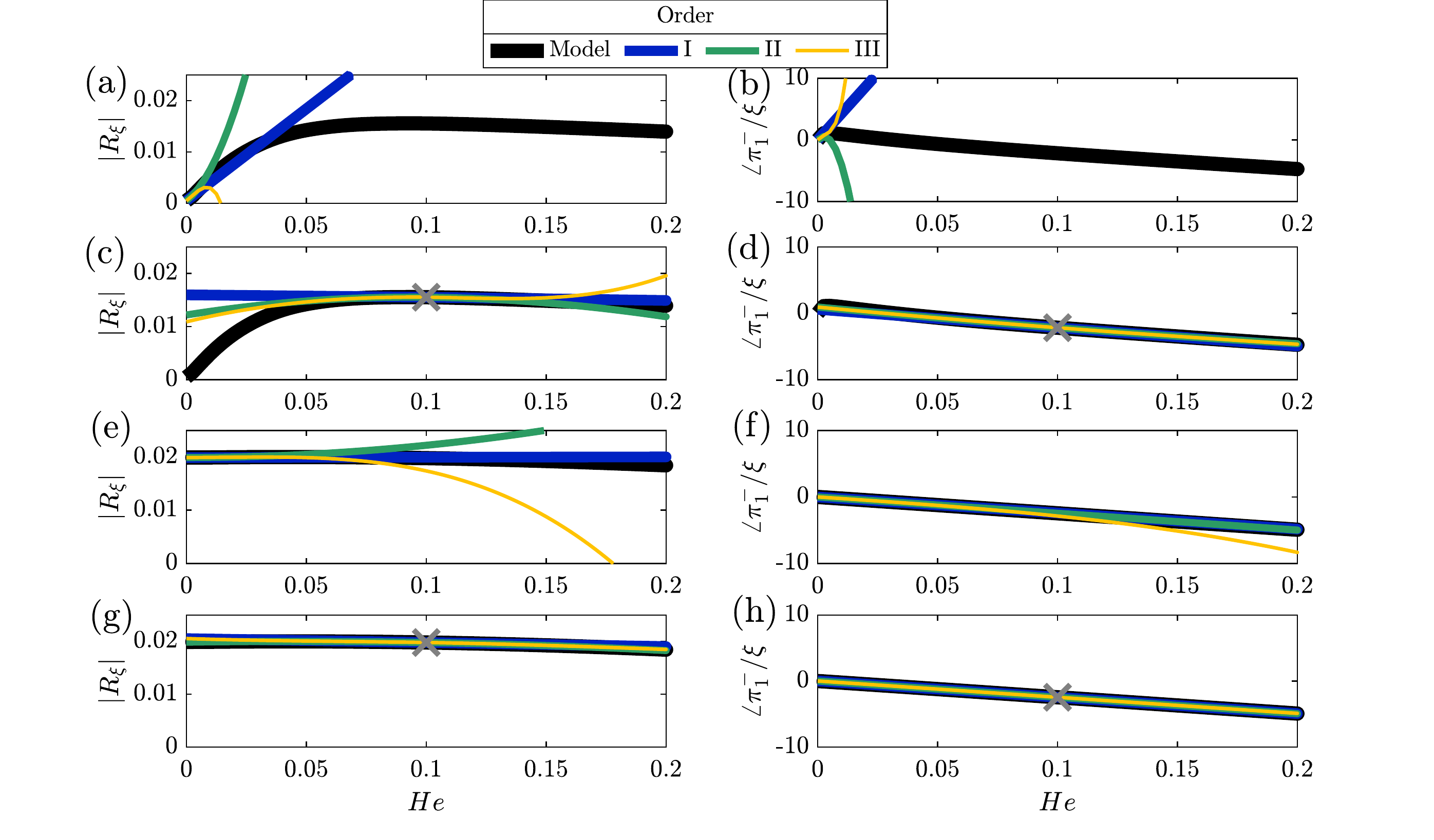}
\caption{Compositional-acoustic reflection coefficient (left) and phase (right) for mixture of air and methane for a subsonic flow in nozzle profile I ($M_1 = M_2 = 0.01$ and $M_t = 0.6$) without dissipation, $f = 0$, (a-d) and with dissipation, $f = 0.08$, (e-h). The colored lines show the predictions using the Taylor expansion. The cross indicates the expansion point. Panels (a,b,e,f) with expansion point at $He_0=0$, panels (c,d,g,h) with expansion point at $He_0=0.1$.  
}
\label{fig:Sensitivity_subsonic}
\end{figure}
\begin{figure}
\centering    
\includegraphics[width=1.0\textwidth]{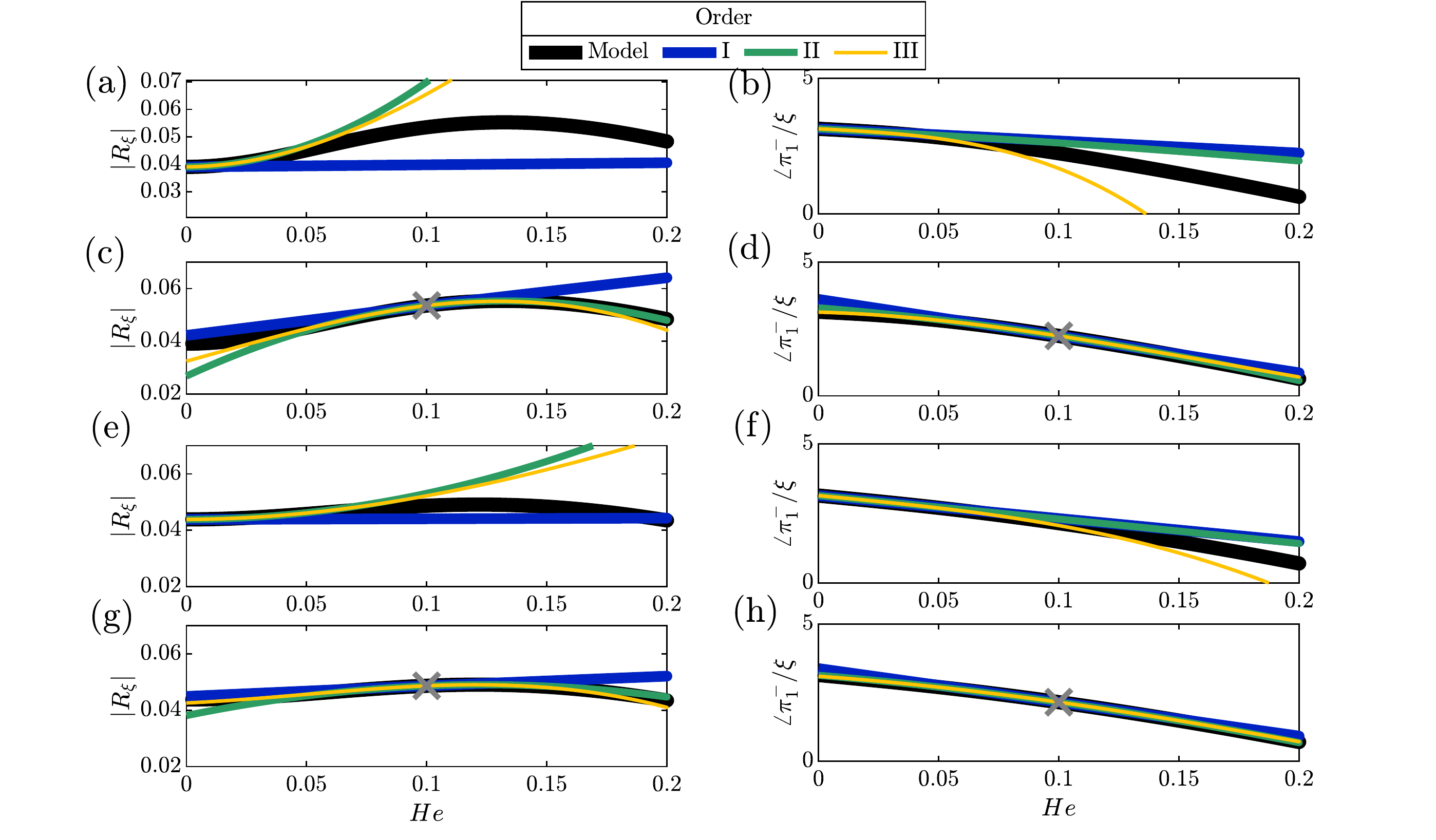}
\caption{{Compositional-acoustic reflection coefficient (left) and phase (right) for mixture of air and methane for a subsonic flow in nozzle profile II ($M_1 = 0.09$, $M_2 = 0.5$, and $M_t = 0.7$) without dissipation, $f = 0$, (a-d) and with dissipation, $f = 0.08$, (e-h). The colored lines show the predictions using the Taylor expansion. The cross indicates the expansion point.  Panels (a,b,e,f) with expansion point at $He_0=0$, panels (c,d,g,h) with expansion point at $He_0=0.1$.  }
}
\label{fig:Sensitivity_subsonic_linvelnozz}
\end{figure}
\begin{figure}
\centering    
\includegraphics[width=1.0\textwidth]{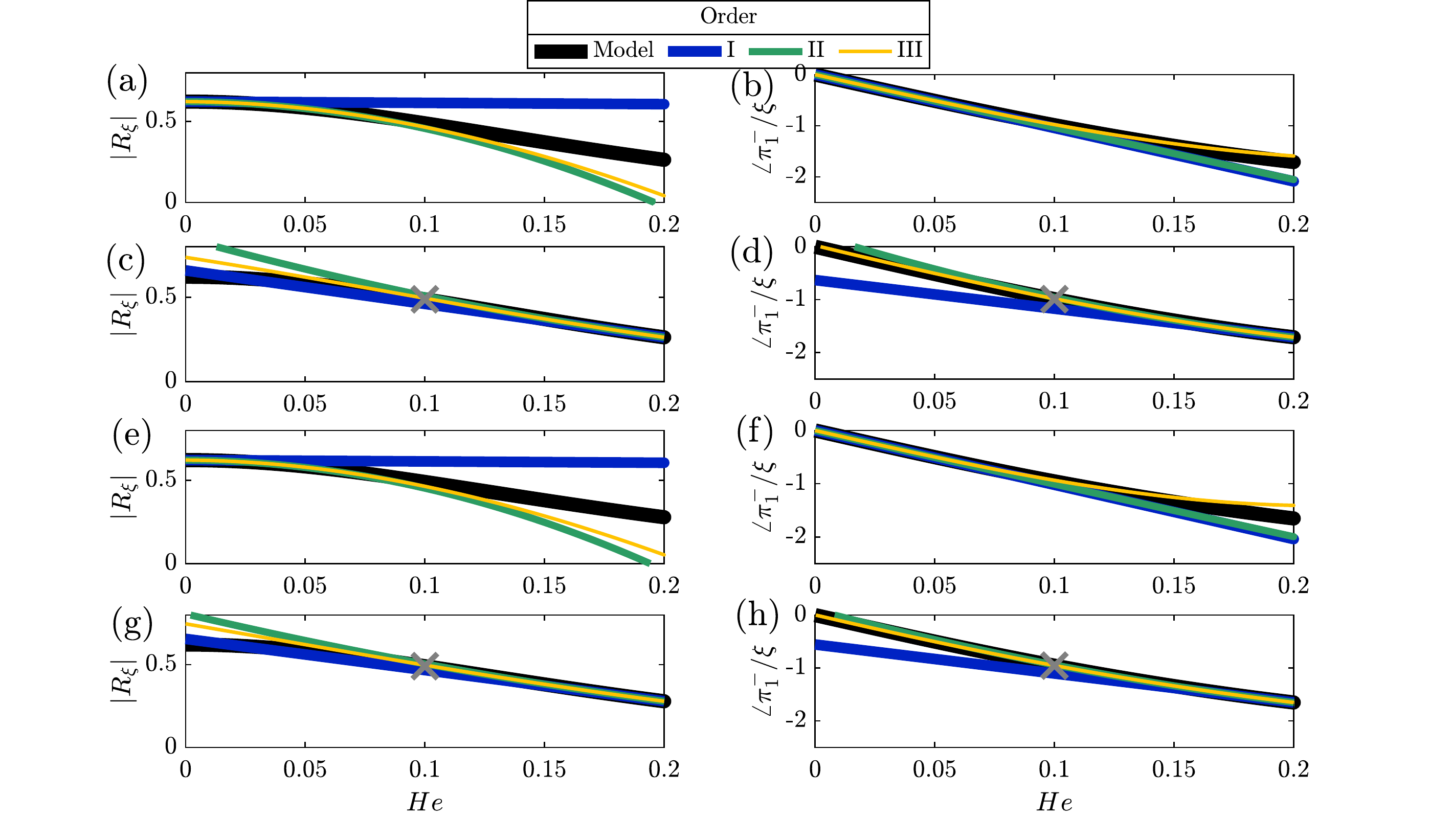}
\caption{Compositional-acoustic reflection coefficient (left) and phase (right) for mixture of air and methane for a supersonic flow without dissipation, $f = 0$, (a-d) and with dissipation, $f = 0.08$, (e-h). The colored lines show the predictions using the Taylor expansion. The cross indicates the expansion point.  Panels (a,b,e,f) with expansion point at $He_0=0$, panels (c,d,g,h) with expansion point at $He_0=0.1$.  }
\label{fig:Sensitivity_supersonic}
\end{figure}
{
Figures \ref{fig:Sensitivity_subsonic} and \ref{fig:Sensitivity_subsonic_linvelnozz} show the contributions of the first three perturbation orders for the subsonic flows, whereas Figure \ref{fig:Sensitivity_supersonic} shows the same quantities for the supersonic flow.
}
The top panels (a,b) show the Taylor expansion at $He_0 = 0$, and the bottom panels (c,d) show the Taylor expansion at $He_0 = 0.1$. \\

On the one hand, in a subsonic flow (Figures~\ref{fig:Sensitivity_subsonic}a,b {and \ref{fig:Sensitivity_subsonic_linvelnozz}a,b}), a third-order expansion around the compact nozzle, $He_0=0$, fails to accurately reproduce both the magnitude and the phase of the model solution. This has a significant consequence on the low-order modelling of nozzles. 
{
Hence, a small change in the Helmholtz number around $He \approx 0$ greatly changes the reflection coefficient.
}
 With the model proposed in this work, we can either calculate the solutions for a range of Helmholtz numbers with no approximation or, alternatively, we can compute one reference solution at $He_0>0$ and expand around it. For example,  if we expand at $He_0=0.1$ (Figures~\ref{fig:Sensitivity_subsonic}c,d {and \ref{fig:Sensitivity_subsonic_linvelnozz}c,d}),  we obtain an accurate approximation of the solution. This is because the solution is less sensitive at $He_0>0$.
The sensitivity at $He = 0$ becomes less significant when the dissipation is modelled, as shown in Figures \ref{fig:Sensitivity_subsonic}e,f {and \ref{fig:Sensitivity_subsonic_linvelnozz}e,f}. 
On the other hand, in a supersonic flow (Figures~\ref{fig:Sensitivity_supersonic}a,b), a third-order expansion around the compact nozzle, $He_0=0$, satisfactorily reproduces both the magnitude and the phase of the solution. The approximation further improves by expanding at $He_0 = 0.1$ (Figures \ref{fig:Sensitivity_supersonic}c,d). Notably, the first-order approximation, which is cheap to compute, accurately quantifies the phase in the supersonic regime for small Helmholtz numbers.

In summary, {the transfer functions at} $He \approx 0$, are  sensitive to small perturbations in the Helmholtz number in the subsonic regime, but less so in the supersonic regime. This sensitivity can be explained on a physical basis. In a subsonic flow in a compact nozzle, {some of} the sound waves generated in the converging section are cancelled by the sound waves generated in the diverging section. However, for larger frequencies ($He > 0$), due to a phase difference induced between the compositional and acoustic waves, the sound waves do not cancel each other. This leads to a sharp rise in the magnitude of the transfer functions near $He \to 0$.
In contrast, in a supersonic flow, the velocity gradient is positive both in the converging and the diverging sections. Hence, the sound waves do not cancel each other, which means that the sensitivity to the Helmholtz number is smaller. 
\FloatBarrier

\end{document}